\documentclass[twocolumn,aps,prl,preprintnumbers]{revtex4}
\usepackage{}
\usepackage{bbm}
\usepackage[latin9]{inputenc}
\usepackage{amsmath}
\usepackage{amssymb}
\usepackage{graphicx}
\usepackage{mathrsfs}
\usepackage{amsfonts}
\usepackage{amsthm}
\usepackage{color}
\usepackage{txfonts}
\usepackage{lipsum}
\usepackage[colorlinks=true,citecolor=blue,linkcolor=blue,urlcolor=blue,anchorcolor=blue]{hyperref}%
\hypersetup{colorlinks=true,citecolor=blue,linkcolor=blue,urlcolor=blue}
\providecommand{\U}[1]{\protect\rule{.1in}{.1in}}
\makeatletter
\@ifundefined{textcolor}{}
{
\definecolor{BLACK}{gray}{0}
\definecolor{WHITE}{gray}{1}
\definecolor{RED}{rgb}{1,0,0}
\definecolor{GREEN}{rgb}{0,1,0}
\definecolor{BLUE}{rgb}{0,0,1}
\definecolor{CYAN}{cmyk}{1,0,0,0}
\definecolor{MAGENTA}{cmyk}{0,1,0,0}
\definecolor{YELLOW}{cmyk}{0,0,1,0}
}
\makeatother
\begin{document}
\title{Higher-order topological solitonic insulators}
\author{Z.-X. Li$^{1}$}
\author{Yunshan Cao$^{1}$}
\author{Peng Yan$^{1}$}
\email[Corresponding author: ]{yan@uestc.edu.cn}
\author{X. R. Wang$^{2,3}$}
\affiliation{$^{1}$School of Electronic Science and Engineering and State Key Laboratory of Electronic Thin Films and Integrated Devices, University of Electronic Science and Technology of China, Chengdu 610054, China}
\affiliation{$^{2}$Physics Department, The Hong Kong University of Science and Technology,
 Clear Water Bay, Kowloon, Hong Kong}
\affiliation{$^{3}$HKUST Shenzhen Research Institute, Shenzhen 518057, China}
\begin{abstract}
Pursuing topological phase and matter in a variety of systems is one central issue in current physical sciences and engineering. Motivated by the recent experimental observation of corner states in acoustic and photonic structures, we theoretically study the dipolar-coupled gyration motion of magnetic solitons on the two-dimensional breathing kagome lattice. We calculate the phase diagram and predict both the Tamm-Shockley edge modes and the second-order corner states when the ratio between alternate lattice constants is greater than a critical value. We show that the emerging corner states are topologically robust against both structure defects and moderate disorders. Micromagnetic simulations are implemented to verify the theoretical predictions with an excellent agreement. Our results pave the way for investigating higher-order topological insulators based on magnetic solitons.
\end{abstract}

\maketitle
\section{INTRODUCTION}
Topological insulators are receiving considerable attention owing to their peculiar properties, such as chiral edge states, and potential applications in spintronics and quantum computing \cite{Hasan2010,Qi2011,Hsieh2009,Pribiag2015}. A conventional $n$-dimensional topological insulator only has $(n-1)$-dimensional (first-order) topological edge/surface modes according to the bulk-boundary correspondence \cite{Hasan2010,Qi2011}. However, a higher-order, e.g., $k$th-order, topological insulator allows $(n-k)$-dimensional topological boundary states with $2\leqslant k\leqslant n$, which goes beyond the standard bulk-boundary correspondence and is characterized by the bulk topological index \cite{Benalcazar2017,Bernevig2017,Song2017,Langbehn2017,Schindler2018,Ezawa2018,Ezawa2018_2,Xie2018}. Interestingly, the experimental evidences of higher-order topological insulators (HOTIs) were reported so far only in classical mechanical and electromagnetic metamaterials \cite{Serra2018,Peterson2018,Xue2019,Ni2019,Hassan2018,Mittal2018,Yang2019,Imhof2018,Serra2019}. In terms of applications of HOTIs in spintronics, it is intriguing to ask if they can exist in magnetic system which is intrinsically, however nonlinear, in contrast to its phononic and photonic counterparts.

Spin waves (or magnons) and magnetic solitons are two important excitations in magnetic system. Various topological states have been predicted in magnonic materials, such as topological magnon insulators \cite{LFZhang2013,Shindou2013,Mook2014,XSWang2017,Chernyshev2016} and magnonic Weyl semimetals \cite{Mook2016,Su2017,Su2017S}. Typical magnetic solitons include vortices \cite{Wachowiak2002, Yamada2007}, bubbles \cite{Bakaul2011,Petit2015}, and skyrmions \cite{Muhlbauer2009,Jiang2015,Yang2018PRB,Yang2018PRL}, which are long-term topics in condensed matter physics for their interesting dynamics and promising applications \cite{Pribiag2007,Finocchio2016}. It has been shown that the collective gyration motion of magnetic solitons exhibits the behavior of waves \cite{Han2013,Yang2017,Mruczkiewicz2016,Behncke2015,Adolff2016}. Recently, the periodic arrangement of ferromagnetic nanodisks, called magnonic crystal, has received much attention in the context of band structure engineering \cite{Behncke2015,Kruglyak2010,Tacchi2011,Krawczyk2014}. When embracing the topological properties of vortex states, it paves the way for robust spintronic information processing. Topological chiral edge states are discovered in a two-dimensional honeycomb lattice of magnetic solitons \cite{Li2018PRB,Kim2017}. However, all these topological magnonic and solitonic states are first-order in nature, according to the classification of topological insulators mentioned above.

\begin{figure}[ptbh]
\begin{centering}
\includegraphics[width=0.48\textwidth]{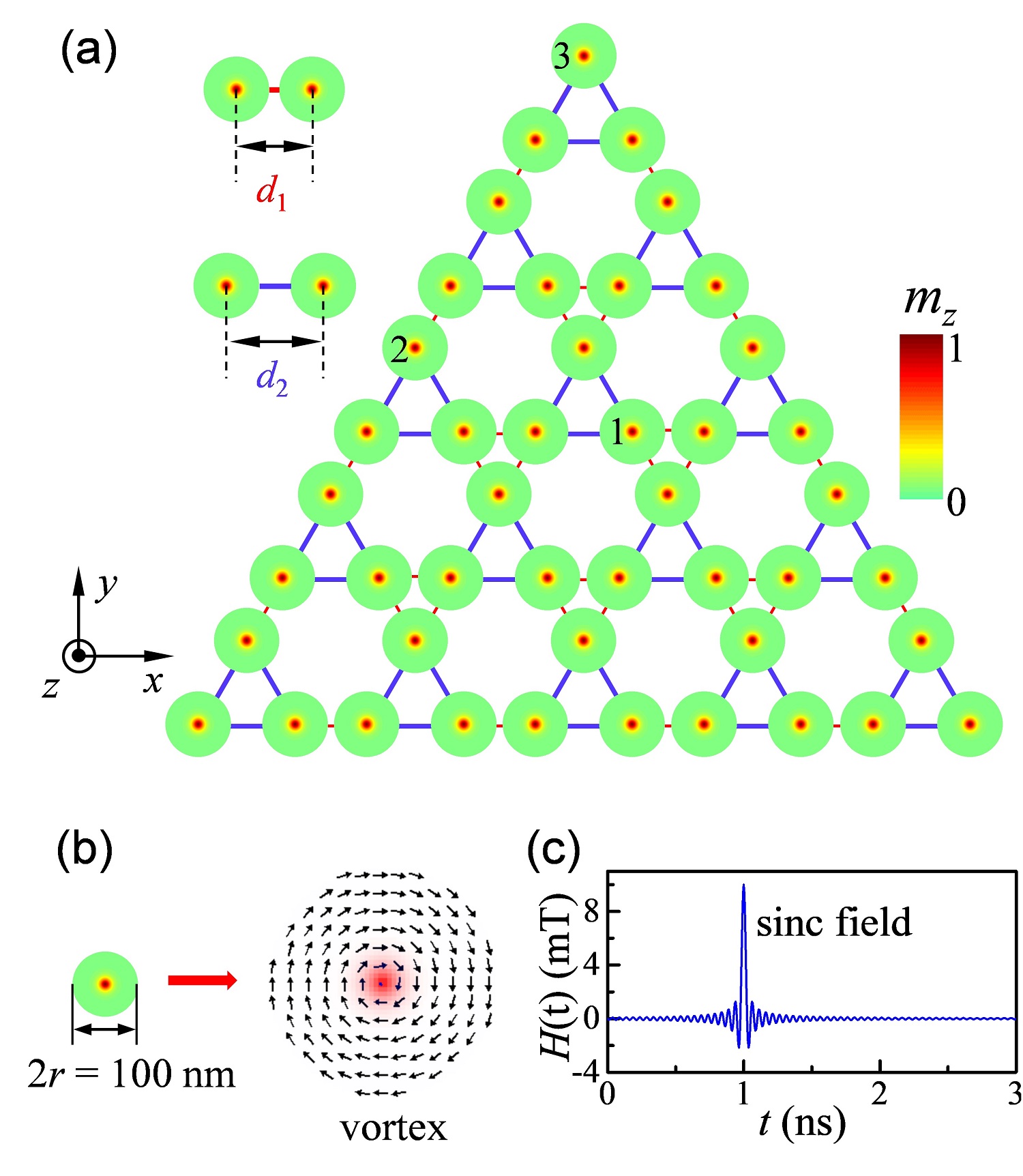}
\par\end{centering}
\caption{\textbf{Triangle-shape breathing kagome lattice of vortices.} (\textbf{a}) Illustration of the triangle-shape breathing kagome lattice including 45 nanodisks of the vortex state. $d_{1}$ and $d_{2}$ are the distance between two nearest-neighbor vortices. Arabic numbers 1, 2 and 3 denote the positions of spectrum analysis for bulk, edge and corner states, respectively. (\textbf{b}) Zoomed in details of a nanodisk with the radius $r=50$ nm and the thickness $w=10$ nm. (\textbf{c}) Time dependence of the sinc-function field $H(t)$ applied to the whole system.}
\label{Figure1}
\end{figure}

In this article, we predict a new class of higher-order topological insulators from the dynamics of magnetic solitons on breathing kagome lattices. Without loss of generality, we use magnetic vortices to demonstrate the principle and as a proof of the concept. The collective motion of vortices is described by the generalized Thiele's equation including an inertial term and a non-Newtonian gyroscopic term. We compute the vortex gyration spectra and find that the system is nontrivial and supports the topological corner states when $d_{2}/d_{1}\geqslant1.2$ ($d_{2}/d_{1}\geqslant1.2$ or $d_{1}/d_{2}\geqslant1.2$) for triangle-shape (parallelogram-shape) lattice. The non-topological Tamm-Shockley edge state is also observed. Here $d_{1}$ and $d_{2}$ are the distances between two kinds of nearest-neighbor vortices, as shown in Fig. \ref{Figure1}(a) and Fig. \ref{Figure4}(a) for triangle and parallelogram structures, respectively. We show that the topological corner states emerge near the gyration frequency of a single vortex, and are robust against structure defects and disorders. Micromagnetic simulations confirm the theoretical results. From an experimental point of view, magnetic soliton lattices can be easily fabricated by electron-beam lithography \cite{Behncke2015}, compared with the complex fabrication processes of phononic and photonic crystals. Our findings shall encourage experimentalists to observe the predicted higher-order topological solitonic states.

\section{Generalized Thiele's equation}
We consider a kagome lattice of magnetic nanodisks with vortex states. Figure \ref{Figure1}(a) plots the lattice structure with alternate distance parameters $d_{1}$ and $d_{2}$. To accurately obtain the gyration spectrum of the vortex array, we start with the generalized Thiele's equation \cite{Li2018PRB}:
\begin{equation}\label{Eq1}
  G_{3}\hat{z}\times\frac{d^{3}\textbf{U}_{j}}{dt^{3}}-M\frac{d^{2}\textbf{U}_{j}}{dt^{2}}+G\hat{z}\times \frac{d\textbf{U}_{j}}{dt}+\textbf{F}_{j}=0,
\end{equation}
where $\mathbf{U}_{j}= \mathbf R_{j} - \mathbf R_{j}^{0}$ is the displacement of the vortex core from the equilibrium position $\mathbf R_{j}^{0}$, $G = -4\pi$$Qw M_{s}$/$\gamma$ is the gyroscopic constant with $Q=\frac{1}{4\pi}\int \!\!\! \int{dxdy\mathbf{m}\cdot(\frac {\partial \mathbf{m}}{\partial {x} } \times \frac {\partial \mathbf{m}}{\partial y } )}$ being the topological charge [$Q=-1/2$ for the vortex configuration shown in Fig. \ref{Figure1}(b)], $\mathbf {m}$ is the unit vector of magnetization, $w$ is the thickness of nanodisk, $M_{s}$ is the saturation magnetization, $\gamma$ is the gyromagnetic ratio, $M$ is the effective mass of the magnetic vortex \cite{Makhfudz2012,Yang2018OE,Buttner2015}, and $G_{3}$ is the third-order gyroscopic coefficient \cite{Mertens1997,Ivanov2010,Cherepov2012}. The conservative force can be expressed as $\textbf{F}_{j}=-\partial W / \partial \mathbf U_{j}$ where $W$ is the potential energy as a function of the vortex displacement: $W=\sum_{j}K\textbf{U}_{j}^{2}/2+\sum_{j\neq k}U_{jk}/2$ with $U_{jk}=I_{\parallel}U_{j}^{\parallel}U_{k}^{\parallel}-I_{\perp}U_{j}^{\perp}U_{k}^{\perp}$ \cite{Kim2017,Shibata2003,Shibata2004}. Here $K$ is the spring constant which is determined by the identity $\omega_{0}= K/|G|$, $\omega_{0}$ is the gyrotropic frequency of a single vortex (see Supplementary Note 1), $I_{\parallel}$ and  $I_{\perp}$ are the longitudinal and transverse coupling constants, respectively.
Imposing $\mathbf{U}_{j}=(u_{j},v_{j})$ and defining $\psi_{j}=u_{j}+i v_{j}$, we have:

\begin{equation}\label{Eq2}
   \hat{\mathcal {D}}\psi_{j}=\omega_{0}\psi_{j}+\sum_{k\in\langle j\rangle}(\zeta^{l}\psi_{k}+\xi^{l} e^{i2\theta_{jk}}\psi^{*}_{k}),
\end{equation}
where the differential operator $\hat{\mathcal {D}}=i\omega_{3}\frac{d^{3}}{dt^{3}}-\omega_{M}\frac{d^{2}}{dt^{2}}+i\frac{d}{dt}$, $\omega_{3}=G_{3}/G$, $\omega_{M}=M/G $, $\zeta^{l}=(I_{\parallel}^{l}-I_{\perp}^{l})/2G $, and $\xi^{l}=(I_{\parallel}^{l}+I_{\perp}^{l})/2G$, with $l=1$ (or $l=2$) representing the distance $d_{1}$ (or $d_{2}$) between nearest neighbor vortices. It is worth mentioning that parameters $I_{\parallel}^{l}$ and $I_{\perp}^{l}$ can be obtained by calculating the eigenfrequencies of a two-vortex system with different combinations of vortex polarities (see Supplementary Note 2). $\theta_{jk}$ is the angle of the direction $\hat{e}_{jk}$ from $x$-axis with $\hat{e}_{jk}=(\mathbf{R}_{k}^{0}-\mathbf{R}_{j}^{0})/|\mathbf{R}_{k}^{0}-\mathbf{R}_{j}^{0}|$ and $\langle j\rangle$ is the set of all intracell and intercell nearest neighbors of $j$.

\begin{figure}[ptbh]
\begin{centering}
\includegraphics[width=0.48\textwidth]{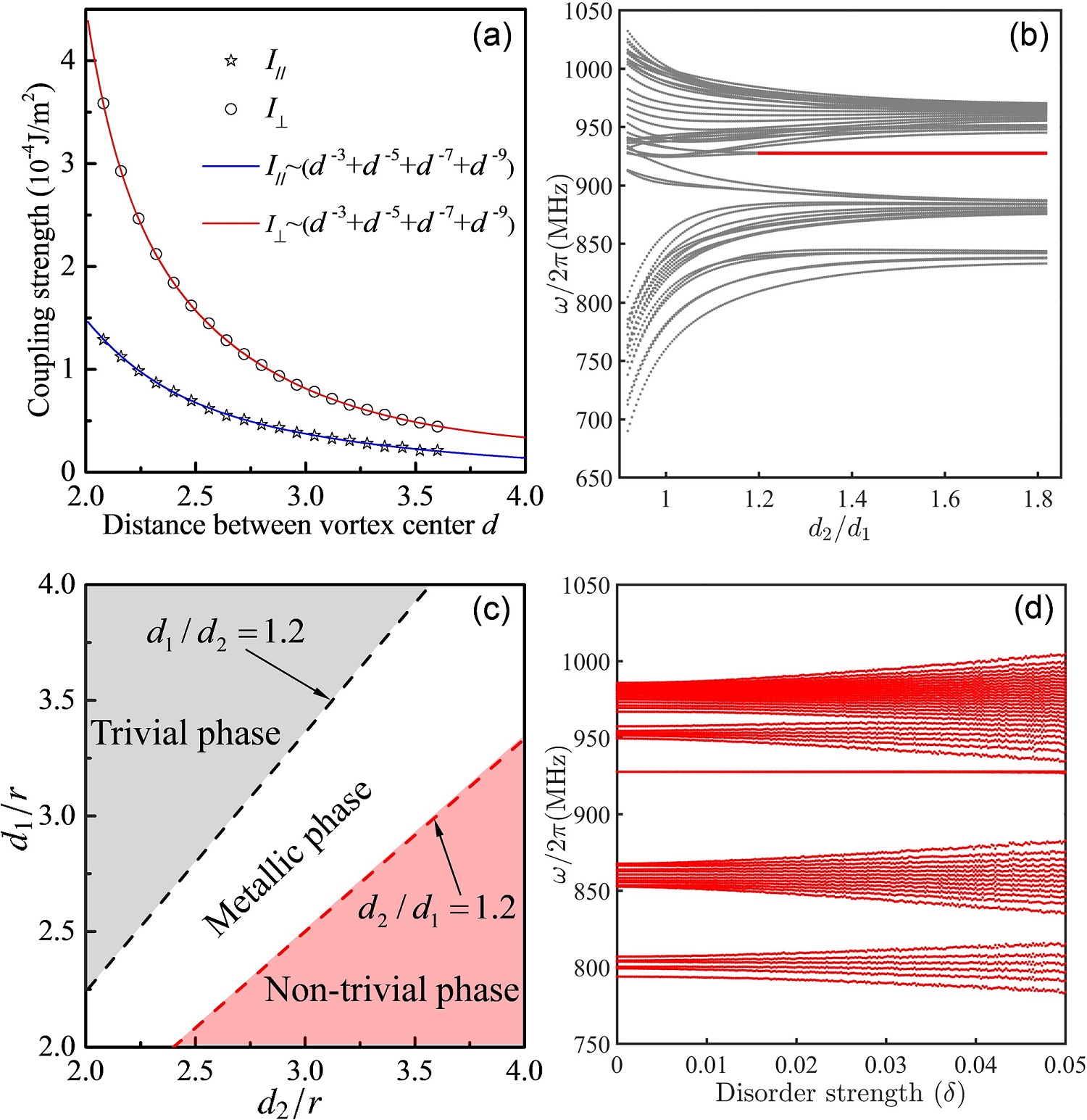}
\par\end{centering}
\caption{\textbf{Corner states and phase diagram for triangle-shape lattice.} (\textbf{a}) Dependence of the coupling strength $I_{\parallel}$ and $I_\perp$ on the vortex-vortex distance $d$ (normalized by the disk radius $r$). Pentagrams and circles denote simulation results and solid curves represent the analytical fitting. (\textbf{b}) Eigenfrequencies of collective vortex gyration under different ratio $d_{2}/d_{1}$ with the red segment labeling the corner state phase. (\textbf{c}) The phase diagram. (\textbf{d}) Eigenfrequencies of the breathing kagome lattice of vortices under different disorder strength.}
\label{Figure2}
\end{figure}

\section{Corner states and phase diagram}

It is known that the coupling strength $I_{\parallel}$ and $I_{\perp}$ strongly depends on the parameter $d$ ($d=d'/r$ with $d'$ the distance between two vortices and $r$ being the radius of nanodisk) \cite{Lee2011,Sukhostavets2013,Sinnecker2014}. Consequently, to calculate the spectrum and the phase diagram of vortex gyrations, we need to know the analytical expression of  $I_{\parallel}(d)$ and $I_{\perp}(d)$. With the help of micromagnetic simulations for a simple system consisting of two nanodisks, we obtain the best fit of the numerical data: $I_{\parallel}=M_{s}^{2}r(-1.72064\times10^{-4}+4.13166\times10^{-2}/d^{3}-0.24639/d^{5}+1.21066/d^{7}-1.81836/d^{9}$) and $I_{\perp}=M_{s}^{2}r(5.43158\times10^{-4}-4.34685\times10^{-2}/d^{3}+1.23778/d^{5}-6.48907/d^{7}+13.6422/d^{9}$), as shown in Fig. \ref{Figure2}(a) with symbols and curves representing simulation results and analytical formulas, respectively. In the calculations, we have adopted the material parameters of Permalloy (Py: Ni$_{80}$Fe$_{20}$) \cite{Yoo2012,Velten2017} with $G=3.0725\times10^{-13}$ Js/m$^{2}$. The spring constant $K$, mass $M$, and non-Newtonian gyration $G_{3}$ are obtained by analyzing the dynamics of a single vortex confined in the nanodisk \cite{Ivanov2010,Ivanov1998}: $K=1.8128\times10^{-3}$ J/m$^{2}$, $M=9.1224\times10^{-25}$ kg, and $G_{3}=4.5571\times10^{-35}$Js$^{3}$/m$^{2}$ (see Supplementary Note 1). Then, by solving Eq. \eqref{Eq2} numerically, we obtain the eigenfrequencies of vortex gyrations in the breathing kagome lattice. Figure \ref{Figure2}(b) shows the eigenfrequencies of the triangle-shape system for different values $d_{2}/d_{1}$ with a fixed $d_{1}=2.2r$. By studying the spatial distribution of the corresponding eigenfunctions, we find that corner states can exist only if $d_{2}/d_{1}\geqslant1.2$, as indicated by the red line segment. Different choices of $d_{1}$ gives almost the same conclusion (see Supplementary Note 3). Furthermore, we calculate the phase diagram by systematically changing $d_{1}$ and $d_{2}$. It is shown that the boundary separating topologically non-trivial and metallic phases lies in $d_{2}/d_{1}=1.2$, while topologically trivial and metallic phases are separated by $d_{1}/d_{2}=1.2$, as shown in Fig. \ref{Figure2}(c). When $d_{2}/d_{1}\geqslant1.2$, the system is topologically non-trivial and can support second-order topological corner states. The system is trivial without any topological edge modes if $d_{1}/d_{2}\geqslant1.2$. Here, the trivial phase is the gapped (insulator) state, the metallic/conducting phase represents the gapless state such that vortices oscillations can propagate in the bulk lattice, and the non-trivial phase means the second-order corner state surviving in a gapped bulk. It is worth mentioning that the critical condition ($d_{2}/d_{1}\geqslant1.2$) for HOTIs may vary with respect to materials parameters. For example, the critical value will slightly increase (decrease) if the radius of the nanodisk increases (decreases).    

Topological corner states should be robust against disorders in the bulk but sensitive to them at corners. To verify these properties of corners states in our system, we calculate the eigenfrequencies of the triangle-shape breathing kagome lattice of vortices under bulk disorders of different strengths (Disorders at three corners are discussed in Supplementary Note 4), as shown in Fig. \ref{Figure2}(d), where $d_{1}=2.08r$ and $d_{2}=3.60r$ ($d_{2}/d_{1}=1.73>1.2$). Here, disorders are introduced by assuming the resonant frequency $\omega_{0}$ with a random shift, i.e., $\omega_{0} \rightarrow \omega_{0}+\delta Z\omega_{0}$, where $\delta$ indicates the strength of the disorder and $Z$ is a uniformly distributed random number between $-1$ to $1$. We average the calculation after 100 realizations of uniformly distributed disorders. Gaussian distribution of $Z$ leads to similar results. We can see from Fig. \ref{Figure2}(d) that with the increasing of the disorder strength, the spectrum of both edge and bulk states is significantly modified, while the corner states are quite robust. Furthermore, the artifact effect of the corner states are also discussed in Supplementary Note 4. These findings echo the observations in photonic and phononic systems \cite{Serra2018,Peterson2018,Xue2019,Ni2019,Hassan2018,Mittal2018,Yang2019,Imhof2018,Serra2019}.
\begin{figure}[ptbh]
\begin{centering}
\includegraphics[width=0.48\textwidth]{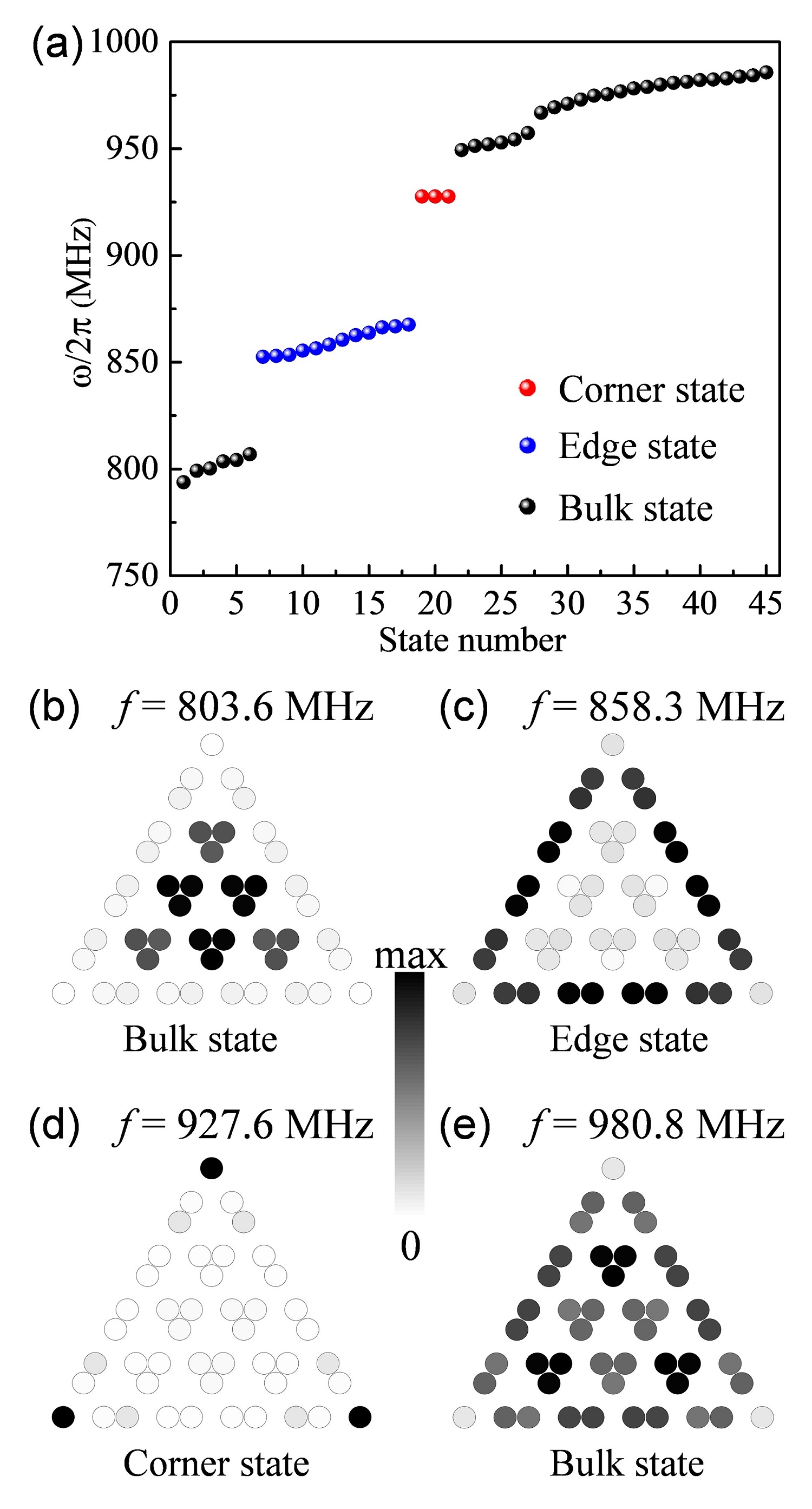}
\par\end{centering}
\caption{\textbf{Eigenmodes in triangle-shape lattice.} (\textbf{a}) Eigenfrequencies of triangle-shape kagome vortex lattice with $d_{1}=2.08r$ and $d_{2}=3.60r$. The spatial distribution of vortex gyrations for the bulk (\textbf{b} and \textbf{e}), edge (\textbf{c}), and corner (\textbf{d}) states.}
\label{Figure3}
\end{figure}

We choose the same geometric parameters as Fig. \ref{Figure2}(d) to explicitly visualize the corner states and other modes in the phase diagram. In this case, from Fig. \ref{Figure2}(a), we have $I_{\parallel}^{1}=1.2894\times10^{-4}$ J/m$^{2}$, $I_{\perp}^{1}=3.5849\times10^{-4}$ J/m$^{2}$, $I_{\parallel}^{2}=2.1237\times10^{-5}$ J/m$^{2}$, and $I_{\perp}^{2}=4.4399\times10^{-5}$ J/m$^{2}$. The computed eigenfrequencies and eigenmodes are plotted in Figs. \ref{Figure3}(a) and \ref{Figure3}(b)-(e). It is found that there are three degenerate modes with the frequency 927.6 MHz, represented by red balls. Then, we confirm that these modes are indeed second-order topological states (corner states) by showing the spatial distribution of vortex gyrations in Fig. \ref{Figure3}(d) with oscillations being highly localized at the three corners. Besides these findings, we also identify the edge states, denoted by blue balls in Fig. \ref{Figure3}(a). The spatial distribution of edge oscillations are confined on three edges, as shown in Fig. \ref{Figure3}(c). However, these edge modes are Tamm-Shockley type \cite{Tamm1932,Shockley1939}, not chiral. They propagate in both directions, that is confirmed in micromagnetic simulations (see Supplementary Note 5). Bulk modes are plotted in Figs. \ref{Figure3}(b) and \ref{Figure3}(e), where corners do not participate in the oscillations.

The higher-order topological properties can be interpreted in terms of the bulk topological index, i.e., the polarization \cite{Smith1993,Vanderbilt1993}:
\begin{equation}\label{Eq3}
 P_{j}=\frac{1}{S}\int\!\!\!\int_{\text{BZ}}A_{j}d^{2}k,
\end{equation}
where $S$ is the area of the first Brillouin zone, $A_{j}=-i\langle\psi|\partial k_{j}|\psi\rangle$ is Berry connection with $j=x,y$, and $\psi$ is the wave function for the lowest band. We have numerically calculated the polarization and find $(P_{x},P_{y})=(0.499,0.288)$ for $d_{1}=2.08r$ and $d_{2}=3.60r$ and $(P_{x},P_{y})=(0.032,0.047)$ for $d_{1}=3r$ and $d_{2}=2.1r$. The former corresponds to the topological insulating phase while the latter is for the trivial phase. Theoretically, the polarization $(P_{x}, P_{y})$ is identical to the Wannier center, which is restricted to two positions for insulating phases. If Wannier center coincides with (0, 0), the system is in the trivial insulating phase and no topological edge states exist. Higher-order topological corner states emerge when the Wannier center lies at (1/2, 1/2$\sqrt{3}$) \cite{Ezawa2018,Xue2019}. The distribution of bulk topological index is consistent with the computed phase diagram Fig. \ref{Figure2}(c).

\begin{figure}[ptbh]
\begin{centering}
\includegraphics[width=0.48\textwidth]{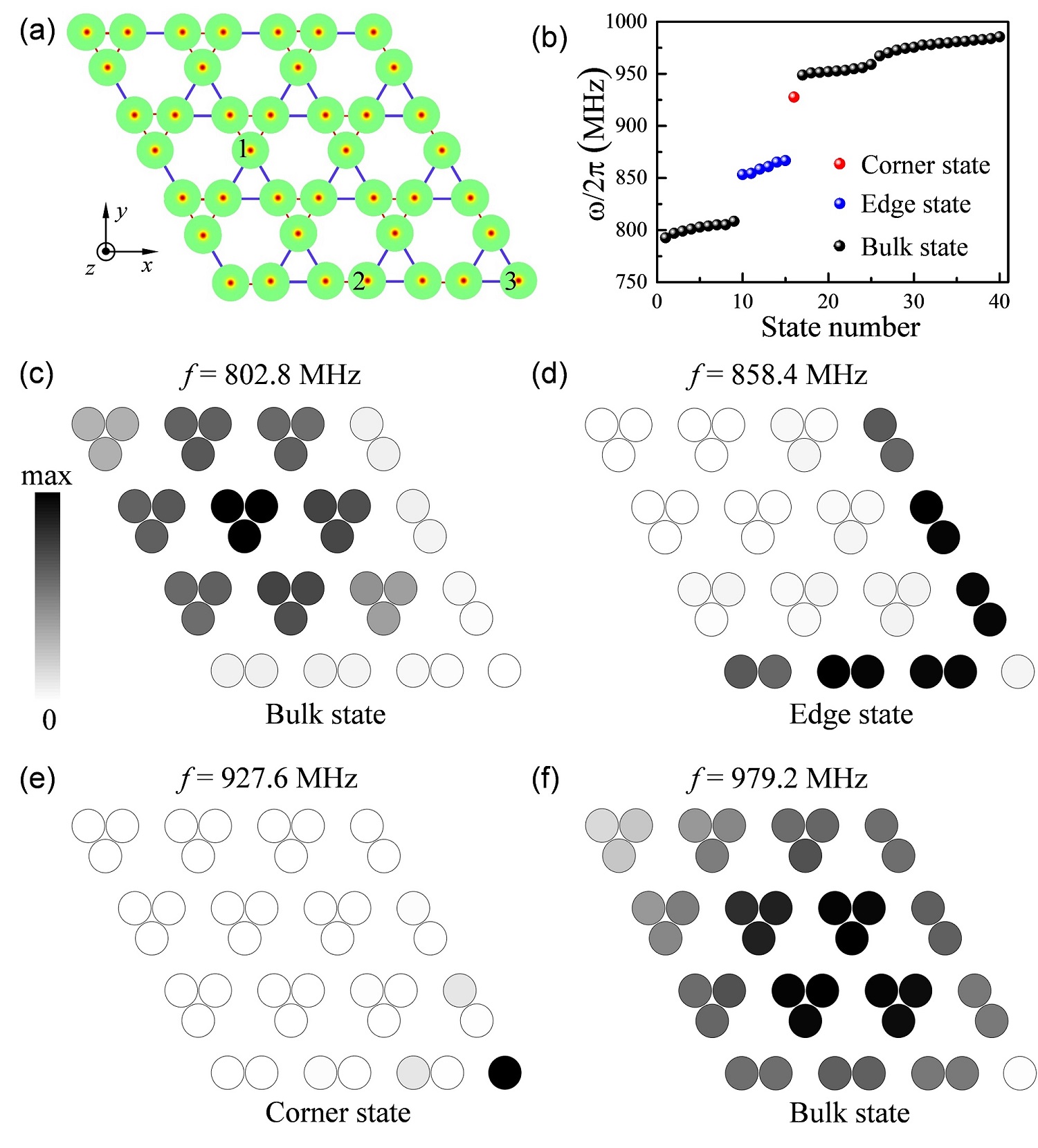}
\par\end{centering}
\caption{\textbf{Eigenmodes in parallelogram-shape lattice.} (\textbf{a}) The sketch for parallelogram-shaped breathing kagome lattice of vortices. Arabic numbers 1, 2 and 3 denote the position of spectrum analysis for bulk, edge, and corner states, respectively. (\textbf{b}) Numerically computed eigenfrequencies for parallelogram-shaped system. The spatial distribution of vortices oscillation for the bulk (\textbf{c} and \textbf{f}), edge (\textbf{d}), and corner (\textbf{e}) states.}
\label{Figure4}
\end{figure}

For completeness, we also study the corner states in another type of breathing kagome lattice of vortices (parallelogram-shape), with the sketch plotted in Fig. \ref{Figure4}(a). We consider the same parameters as those in the triangle-shape lattice. Figure \ref{Figure4}(b) shows the eigenfrequencies of system which are obtained by numerically solving Eq. (2). Interestingly, we see that there is only one corner state at the frequency equal to 927.6 MHz, represented by the red ball. Edge and bulk states are also observed, denoted by blue and black balls, respectively. To have a better understanding of these modes, we have plotted the spatial distribution of vortices oscillation, as shown in Figs. \ref{Figure4}(c)-\ref{Figure4}(f). From Fig. \ref{Figure4}(e), one can clearly see that the oscillations for corner state are confined to one acute angle and the vortex at the position of two obtuse angles hardly oscillates. The spatial distribution of vortex gyration for edge and bulk states are plotted in Figs. \ref{Figure4}(c), \ref{Figure4}(d), and \ref{Figure4}(f). The robustness of the corner states and the phase diagram are discussed in Supplementary Note 6.\\

It is interesting to note that the results of triangle-shape and parallelogram-shape lattices are closely related. Two opposite acuted-angle corners in the parallelogram are actually not equivalent: one via $d_{1}$ bonding while the other one via $d_{2}$ bonding; see Fig. 4(a). Only the $d_{2}$ bonding (bottom-right) corner in the parallelogram-shape lattice is identical to three corners in the triangle-shape lattice. Therefore, for parallelogram-shape lattice, we can observe only one corner state either in the bottom-right corner [when $d_{2}/d_{1}\geqslant1.2$; see Fig. \ref{Figure4}(f) in the main text] or in the top-left corner [when $d_{1}/d_{2}\geqslant1.2$; see Supplementary Figure 7].
\begin{figure}[ptbh]
\begin{centering}
\includegraphics[width=0.48\textwidth]{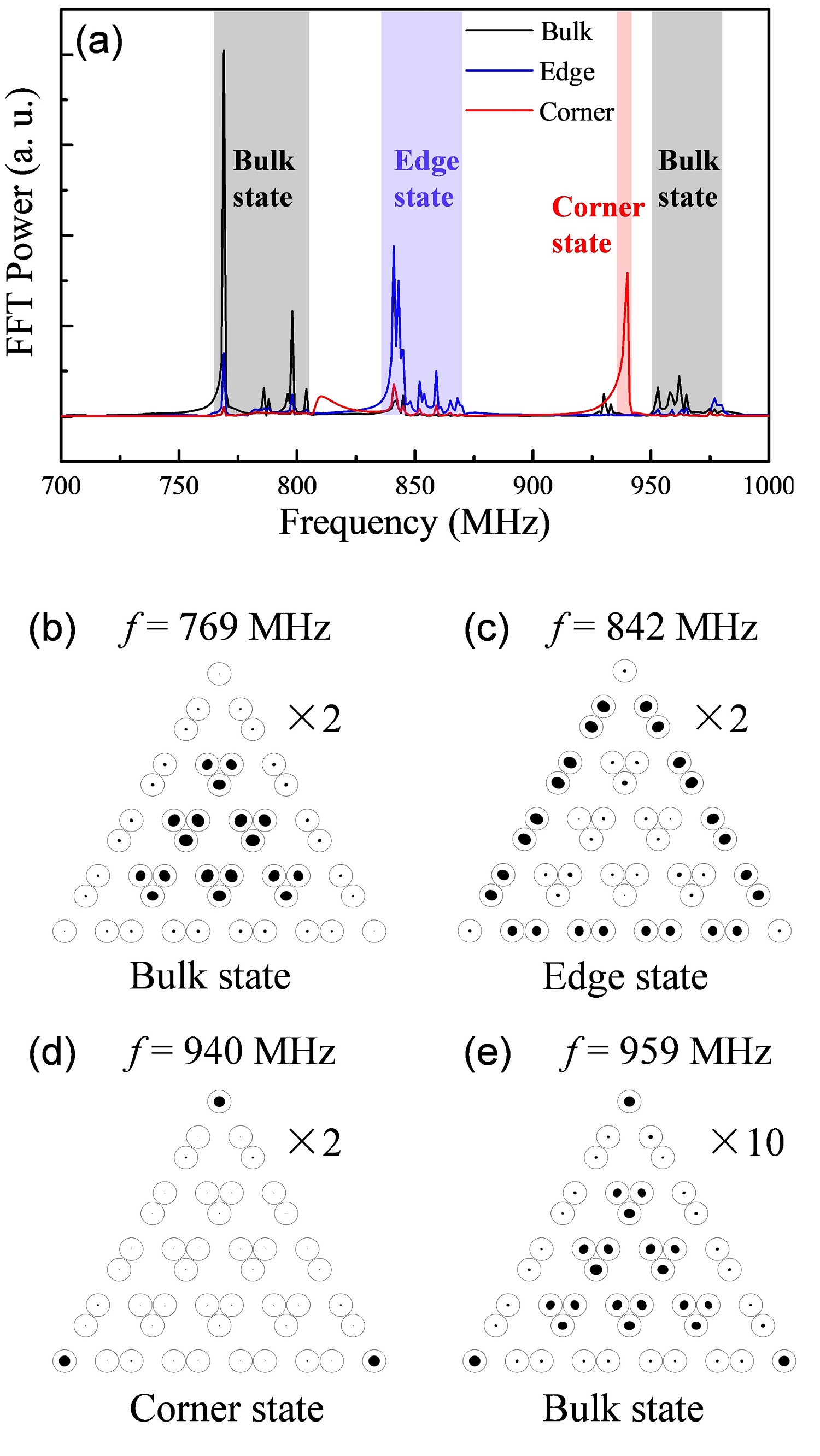}
\par\end{centering}
\caption{\textbf{Micromagnetic simulation of excitations in triangle-shape structure.} (\textbf{a}) The temporal Fourier spectrum of the vortex oscillations at different positions. The spatial distribution of oscillation amplitude under the exciting field of various frequencies, 769 MHz (\textbf{b}), 842 MHz (\textbf{c}), 940 MHz (\textbf{d}), and 959 MHz (\textbf{e}). Since the oscillation amplitudes of the vortex centers are too small, we have magnified them by 2 or 10 times labeled in each figure.}
\label{Figure5}
\end{figure}

\begin{figure}[ptbh]
\begin{centering}
\includegraphics[width=0.48\textwidth]{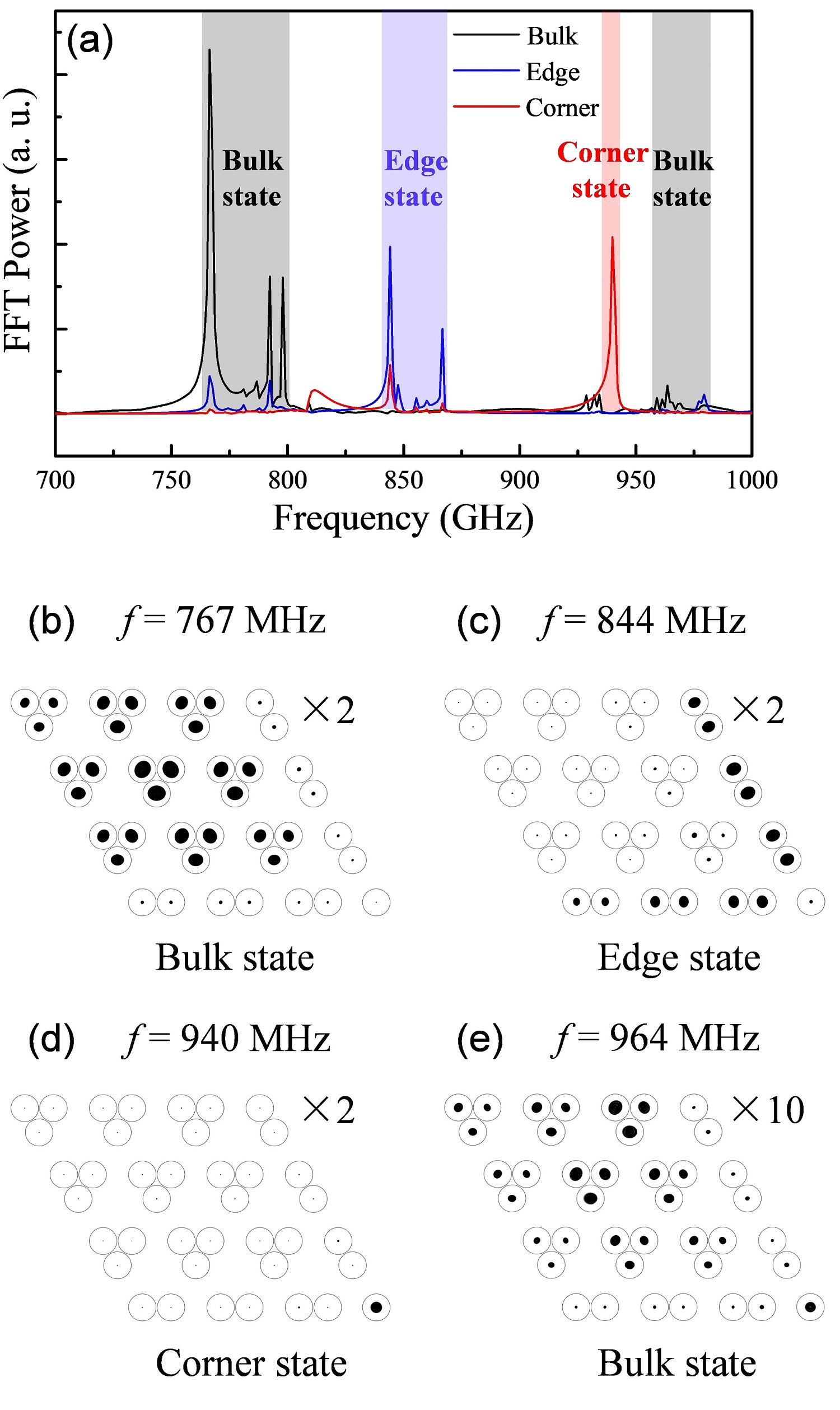}
\par\end{centering}
\caption{\textbf{Micromagnetic simulation of excitations in parallelogram-shape structure.} (\textbf{a}) The temporal Fourier spectrum of the vortex oscillations at different positions. The spatial distribution of oscillation amplitude under the exciting field with different frequencies, 767 MHz (\textbf{b}), 844 MHz (\textbf{c}), 940 MHz (\textbf{d}), and 964 MHz (\textbf{e}). The simulation time is 100 ns. Since the oscillation amplitudes of the vortices centers are too small, we have magnified them by 2 or 10 times labeled in each figure.}
\label{Figure6}
\end{figure}

\section{Micromagnetic simulations}
To verify the theoretical predictions of HOTIs above, we perform full micromagnetic simulations by considering a breathing kagome lattice consisting of a few identical magnetic nanodisks in vortex states, as shown in Fig. \ref{Figure1}(a) and Fig. \ref{Figure4}(a), with the same geometric parameters as those in Fig. \ref{Figure3} and Fig. \ref{Figure4}, respectively. Micromagnetic software MUMAX3 \cite{Vansteenkiste2014} is used to simulate the dynamics of vortices in Py. The material parameters are as follows: the saturation magnetization $M_{s}=0.86\times10^{6}$ A/m, the exchange stiffness $A=1.3\times10^{-11}$ J/m, and the Gilbert damping constant $\alpha=10^{-4}$ (in order to observe the vortex oscillations clearly, we have chosen a rather small damping parameter). In the simulations, we set the cell size to be $2\times2\times10 $ nm$^{3}$. To excite the full spectrum (up to a cut-off frequency) of the vortex oscillations, we apply a sinc-function magnetic field $H(t)=H_{0}\sin[2\pi f(t-t_{0})]/[2\pi f(t-t_{0})]$ along the $x$-direction with $H_{0}=10$ mT, $f=20$ GHz, and $t_{0}=1$ ns, as plotted in Fig. \ref{Figure1}(c). The exciting field is applied over the whole system. The spatiotemporal evolutions of the vortices center $\textbf{R}_{j}=(R_{j,x}, R_{j,y}$) in all nanodisks are recorded every 200 ps, with the total simulation time being 1000 ns. Here $R_{j,x}$ and $R_{j,y}$ are defined by $R_{j,x}=\frac {\int \!\!\! \int{x|m_{z}|^{2}dxdy}}{\int \!\!\! \int{|m_{z}|^{2}dxdy}}$ and $R_{j,y}=\frac {\int \!\!\! \int{y|m_{z}|^{2}dxdy}}{\int \!\!\! \int{|m_{z}|^{2}dxdy}}$, with the integral region being confined in the $j$-th nanodisk.

To identify the energy band of higher-order topological edge states in triangle-shape lattice, we compute the temporal Fourier spectrum of the vortex oscillations at different positions [labeled with arabic numbers 1, 2 and 3, see Fig. \ref{Figure1}(a)]. Figure \ref{Figure5}(a) shows the spectra, with black, blue, and red curves denoting the positions of bulk (Number 1), edge (Number 2) and corner (Number 3) bands, respectively. One can immediately see that, near the frequency of 940 MHz, the spectrum for the corner has a very strong peak, which does not happen for the edge and bulk. We therefore infer that this is the corner-state band with oscillations localized only at three corners. Similarly, one can identify the frequency range which allows the bulk and edge states, as shown by shaded area with different colors in Fig. \ref{Figure5}(a). To visualize the spatial distribution of vortex oscillations for different modes, we choose four representative frequencies: 940 MHz for the corner state, 842 MHz for the edge state, and both 769 MHz and 959 MHz for bulk states, and then stimulate their dynamics by a sinusoidal field $\textbf{h}(t)=h_0\sin(2\pi ft)\hat{x}$ with $h_0=0.1$ mT applied to the whole system for 100 ns. We plot the spatial distribution of oscillation amplitude in Figs. \ref{Figure5}(b)-\ref{Figure5}(e). One can clearly see the corner state in Fig. \ref{Figure5}(d), which is in a good agreement with theoretical results shown in Fig. \ref{Figure3}(d) (theoretically calculated corner state locates at 927.6 MHz). Spatial distribution of vortices motion for bulk and edge states are shown in Figs. \ref{Figure5}(b) and \ref{Figure5}(c), respectively. It is noted that vortices at three corners in Fig. \ref{Figure5}(e) also oscillate with a sizable amplitude, which is somewhat quite unexpected for bulk states. We attribute this inconsistency to the strong coupling (or hybridization) between the bulk and corner modes, since their frequencies are very close to each other, as shown in Figs. \ref{Figure3}(a) and \ref{Figure5}(a).

Like the triangle-shaped case, we have identified the corner, edge, and bulk states by micromagnetic simulation in parallelogram-shaped lattice with the same sinusoidal exciting fields applied to the whole system. We compute the temporal Fourier spectrum of the vortex oscillations at different positions [denoted with arabic numbers 1, 2 and 3, see Fig. \ref{Figure4}(a)]. The spectra are shown in Fig. \ref{Figure6}(a) with the black, blue and red curves indicating the positions of bulk (Number 1), edge (Number 2) and corner (Number 3) bands, respectively. Shaded area with different colors denote different modes. The spatial distribution of oscillation amplitude is plotted in Figs. \ref{Figure6}(b)-\ref{Figure6}(e). Fig. \ref{Figure6}(d) shows only one corner state at only one (bottom-right) acute angle, which is in a good agreement with theoretical results shown in Fig. \ref{Figure4}(e). Spatial distribution of vortices motion for bulk and edge states are shown in Figs. \ref{Figure6}(b) and \ref{Figure6}(c), respectively. Interestingly, the hybridization between the bulk mode and corner mode occurs as well in parallelogram-shaped breathing kagome lattice, see Fig. \ref{Figure6}(e).

\section{CONCLUSIONS}
We have investigated the higher-order topological insulator in triangle-shaped and parallelogram-shaped breathing kagome lattice of magnetic vortices. Phase diagram including various solitonic states was obtained theoretically. It was found that the second-order topological corner state emerge only under a critical geometric parameter. We interpreted these results by the bulk topological index. Micromagnetic simulations were performed to confirm all theoretical predictions. We envision the existence of higher-order topological solitonic insulators in other type of lattices (breathing honeycomb, for instance), which is an interesting issue for future study. Identifying higher-order topological magnon insulator is also an open question. We believe that the findings presented in this work shall encourage experimentalists to find higher-order topological states in magnetic systems, within current technology reach.

\begin{acknowledgments}
%\section*{ACKNOWLEDGMENTS}
This work was supported by the National Natural Science Foundation of China (NSFC) (Grants No. 11604041 and 11704060), the National Key Research Development Program under Contract No. 2016YFA0300801, and the National Thousand-Young-Talent Program of China. X.R.W. was supported by Hong Kong RGC (Grants No. 16300117, 16301518 and 16301619). Z.-X. L. acknowledged financial support of NSFC under Grant No. 11904048. We acknowledged C.W. and X.S.W. for helpful discussions.
\end{acknowledgments}

\newpage
\onecolumngrid
\quad\par
\centerline{\bf{Supplementary Information}}
%\quad\par
%\centerline{\bf{Higher-order topological solitonic insulators}}
%\quad\par
%\centerline{Z.-X. Li, Yunshan Cao, Peng Yan, and X. R. Wang}
\quad\par
{
\noindent We provide here information about the determination of the spring constant $K$, the vortex mass $M$,
and the non-Newtonian gyration coefficient $G_{3}$ in Supplementary Note 1, the calculation of coupling parameters $I_{\parallel}$ and $I_{\perp}$ between two vortices in Supplementary Note 2, and the eigenfrequencies of triangle-shape breathing kagome vortex lattice for different lattice constants in Supplementary Note 3. For triangle-shape lattice, the numerical demonstration of the robustness of corner states against disorders and defects are provided in Supplementary Note 4. The Tamm-Shockley edge state is shown in Supplementary Note 5. Finally the robustness of the corner states and the phase diagram for parallelogram-shape lattice are discussed in Supplementary Note 6.}

\twocolumngrid
\maketitle
%\quad\par

%\quad\par
\subsection*{\bf{Supplementary Note 1}}
\noindent We determine the spring constant $K$, vortex mass $M$, and non-Newtonian gyration $G_{3}$ through the following relations \cite{Ivanov2010,Ivanov1998}: $\omega_{0}= K/G$, $G_{3}\bar{\omega}^{2}=G$, $G_{3}(\Delta\omega+\omega_{0})=M$, $2\bar{\omega}=\omega_{1}+\omega_{2}$, and $\Delta\omega=|\omega_{2}-\omega_{1}|$, where $\omega_{0}$ is the frequency of the gyroscopic mode, $\omega_{1}$ and $\omega_{2}$ are the frequencies of the other two higher-order modes with opposite gyration handedness \cite{Ivanov2010}. For Py, $G=3.0725\times10^{-13}$ Js/m$^{2}$. We first numerically obtain the excitation spectrum of an isolated nanodisk with a vortex, with results plotted in Supplementary Fig. \ref{FigureS1}. We confirm that the lowest mode is an anti-clockwise gyrotropic mode and the other two high-frequency modes have opposite directions of gyration \cite{Yoo2015}. The frequency of the three peaks reads $\omega_{0}=2\pi\times0.939$ GHz, $\omega_{1}=2\pi\times11.945$ GHz, and $\omega_{2}=2\pi\times14.192$ GHz. Substituting into the formula above, we get $K=1.8128\times10^{-3}$ J/m$^{2}$, $M=9.1224\times10^{-25}$ kg, and $G_{3}=4.5571\times10^{-35}$Js$^{3}$/m$^{2}$.

\begin{figure}[ptbh]
\begin{centering}
\includegraphics[width=0.42\textwidth]{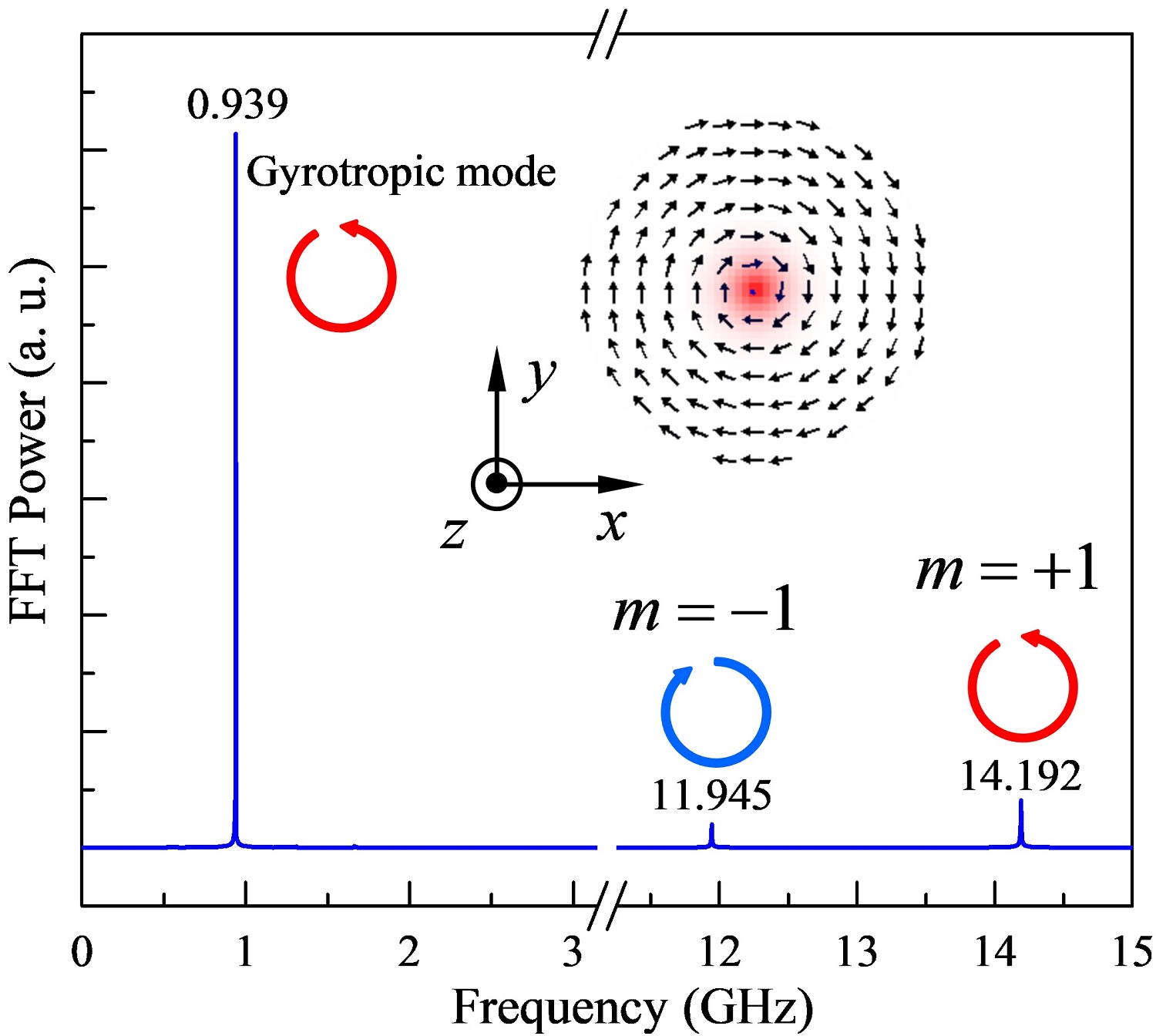}
\par\end{centering}
\caption{\textbf{Spectrum of a single vortex in a nanodisk:}
Inset shows the micromagnetic structure of the nanodisk. The external pulse field with strength $h_{0}=10$ mT and duration time $t=10$ ps is applied along the $x$-direction. Arrows denote the handedness of the vortex gyration at each mode.}
\label{FigureS1}
\end{figure}

\begin{figure}[ptbh]
\begin{centering}
\includegraphics[width=0.48\textwidth]{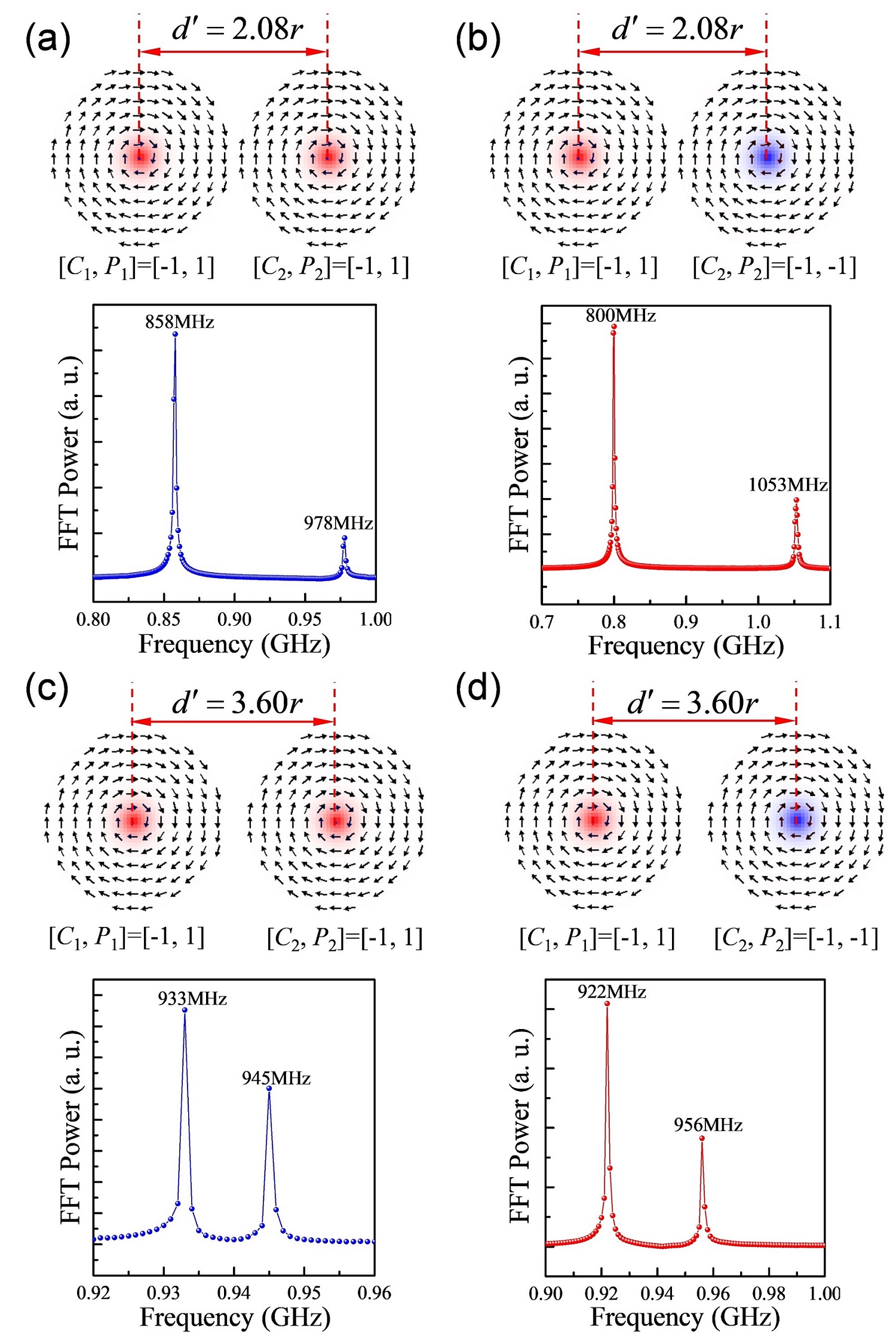}
\par\end{centering}
\caption{\textbf{Dynamics of vortex-vortex system:} Micromagnetic structures and spectra of the vortex-vortex system with different combinations of vortex polarities under $d'=2.08r$ [(\textbf{a}) and (\textbf{b})] and $d'=3.60r$ [(\textbf{c}) and (\textbf{d})]. $C_{i}$ and $P_{i}$ ($i=1$ or 2) represent the chirality and the polarity of vortices, respectively.}
\label{FigureS2}
\end{figure}

\subsection*{\bf{Supplementary Note 2}}
\noindent To determine the two key parameters $I_{\parallel}$ and $I_{\perp}$ in the formula $W=\sum_{j}K\textbf{U}_{j}^{2}/2+\sum_{j\neq k}U_{jk}/2$ with $U_{jk}=I_{\parallel}U_{j}^{\parallel}U_{k}^{\parallel}-I_{\perp}U_{j}^{\perp}U_{k}^{\perp}$, $U_{j}^{\parallel}=\textbf{U}_{j}\cdot\hat{e}_{jk}$, and $U_{j}^{\perp}=\textbf{U}_{j}\cdot(\hat{z}\times\hat{e}_{jk})$, we consider a simplified system consisting of two nanodisks (with the same size as those in the main text), as shown in Supplementary Fig. \ref{FigureS2}. The eigenfrequencies of coupled two-vortex system can be expressed as $\omega=\omega_{0}\sqrt{(1\pm I_{\parallel}/K)(1\mp P_{1}P_{2}I_{\perp}/K)}$ \cite{Lee2011}, where $P_{1}$ (or $P_{2}$) is either $+1$ or $-1$ depending on the vortex polarity. Therefore, once the frequencies of coupled modes for different combinations of vortex polarities ($P_{1}P_{2}=1$ or $P_{1}P_{2}=-1$) are determined from micromagnetic simulation, we can derive $I_{\parallel}$ and $I_{\perp}$ according to the dispersion relation. Supplementary Figs. \ref{FigureS2}(a) and \ref{FigureS2}(b) [Supplementary Figs. \ref{FigureS2}(c) and \ref{FigureS2}(d)] show the setup and spectra of the two-vortex system with $d'=2.08r$ ($d'=3.60r$). By extracting the frequency corresponding to the peak and after some algebra, we obtain $I_{\parallel}^{1}=1.2894\times10^{-4}$ J/m$^{2}$, $I_{\perp}^{1}=3.5849\times10^{-4}$ J/m$^{2}$, $I_{\parallel}^{2}=2.1237\times10^{-5}$ J/m$^{2}$, and $I_{\perp}^{2}=4.4399\times10^{-5}$ J/m$^{2}$. Further, we have systematically calculated the $d-$dependence of coupling parameters $I_{\parallel}$ and $I_{\perp}$, as shown in Fig. 2(a) in the main text.

\begin{figure}[ptbh]
\begin{centering}
\includegraphics[width=0.48\textwidth]{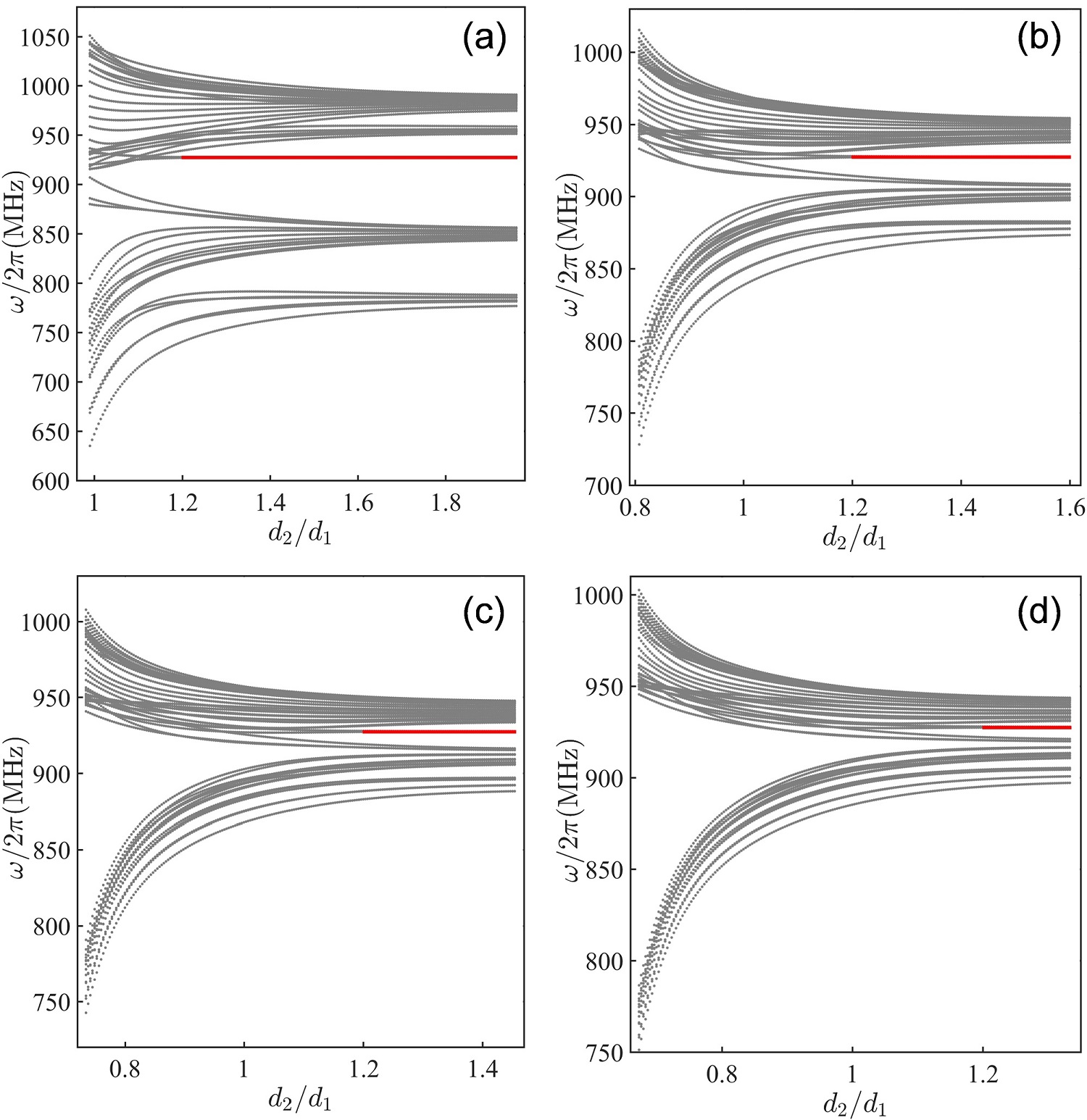}
\par\end{centering}
\caption{\textbf{Topological phase diagram of triangle-shape breathing kagome lattice of vortices:} Eigenfrequencies of system for different values $d_{2}/d_{1}$ with $d_{1}=2.04r$ (\textbf{a}), $d_{1}=2.5r$ (\textbf{b}), $d_{1}=2.75r$ (\textbf{c}), and $d_{1}=3r$ (\textbf{d}). The red line segment represents the region where the corner states exist.}
\label{FigureS3}
\end{figure}

\begin{figure}[ptbh]
\begin{centering}
\includegraphics[width=0.48\textwidth]{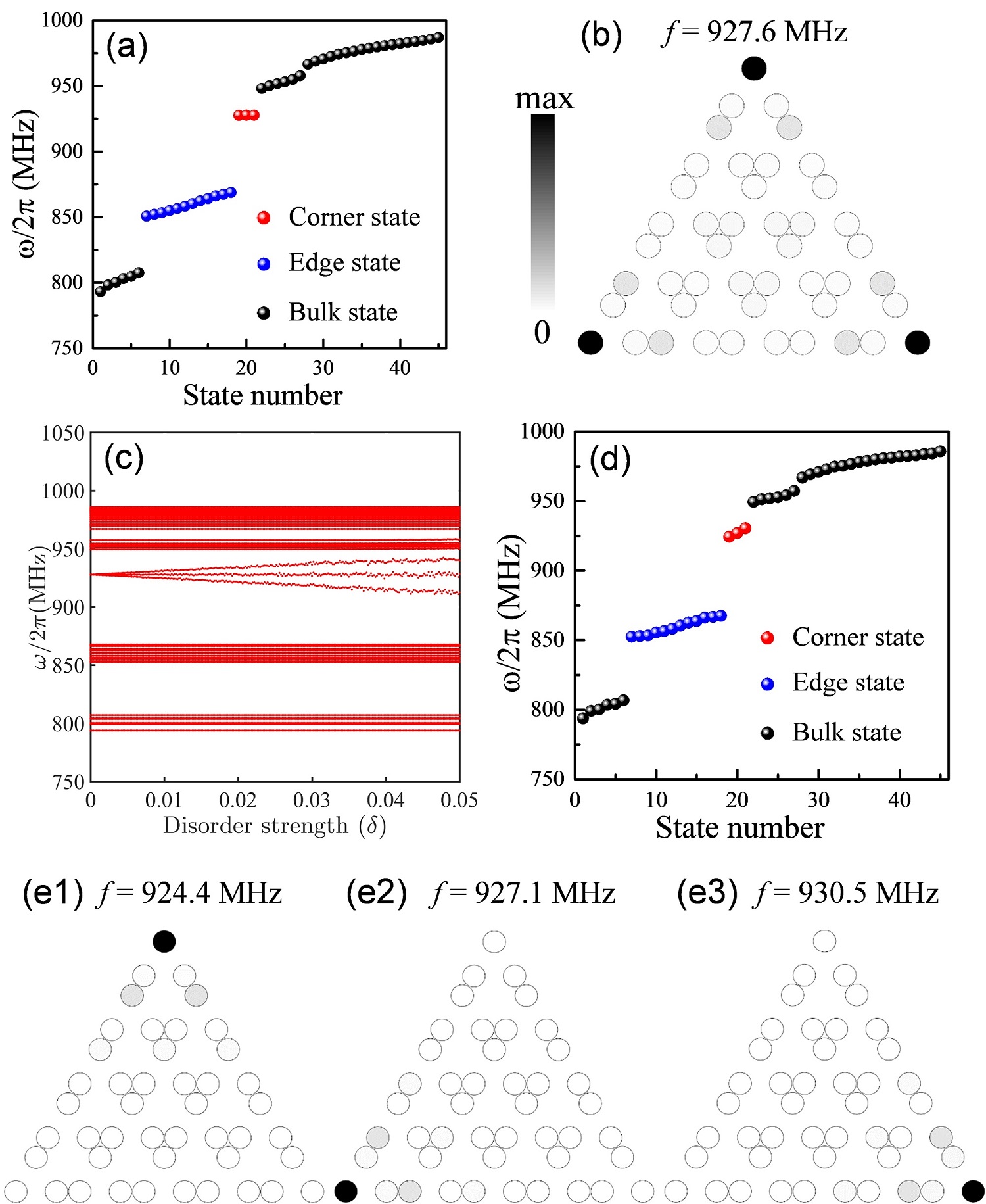}
\par\end{centering}
\caption{\textbf{Disorder effect of the corner states:} A case study with $\delta=0.01$ for disorders (\textbf{a}) in the bulk and (\textbf{d}) at corners. (\textbf{c}) Eigenfrequencies of the triangle-shape breathing kagome lattice of vortices under different disorder strengths where the disorders at three corners. (\textbf{b}) and (\textbf{e1})-(\textbf{e3}) eigenmodes of corner states as shown in (\textbf{a}) and (\textbf{d}), respectively.}
\label{FigureS4}
\end{figure}

\subsection*{\bf{Supplementary Note 3}}
\noindent As mentioned in the main text, for triangle-shape breathing kagome lattice of vortices, the phases of the system are entirely determined by the ratio $d_{2}/d_{1}$. To further confirm this conclusion, we plot the eigenfrequencies of system for different value $d_{2}/d_{1}$, with four choices of $d_{1}$, as shown in Supplementary Figs. \ref{FigureS3}(a)-\ref{FigureS3}(d). It can be clearly seen that regardless of the value of $d_{1}$, the system always shows corner states when $d_{2}/d_{1}\geqslant1.2$.

\begin{figure}[ptbh]
\begin{centering}
\includegraphics[width=0.48\textwidth]{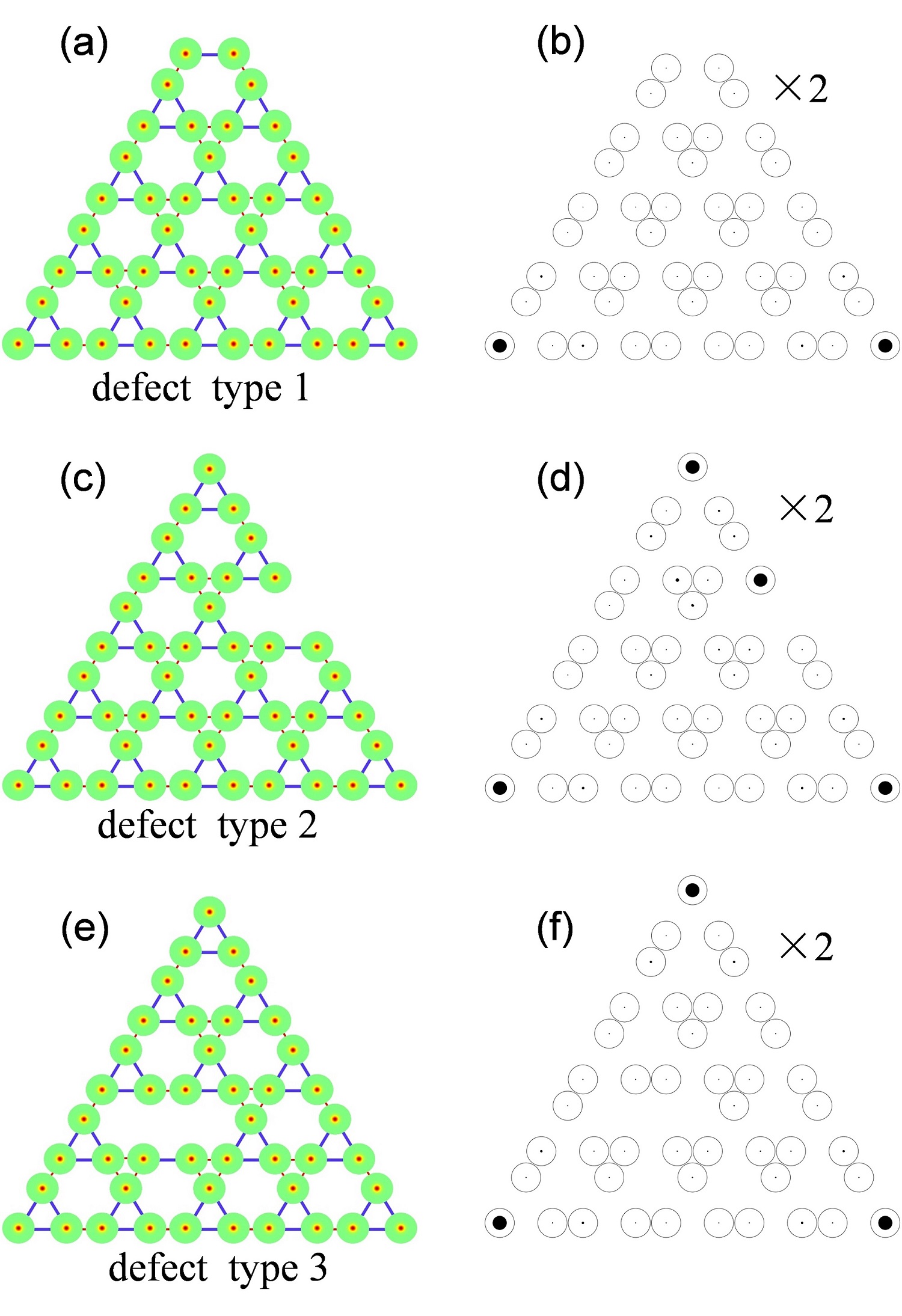}
\par\end{centering}
\caption{\textbf{Artifact effect of the corner states:} The schemes and spatial distribution of oscillation amplitude for different type of defects, which locates at the corner [(\textbf{a}) and (\textbf{b})], the edge [(\textbf{c}) and (\textbf{d})], and the center [(\textbf{e}) and (\textbf{f})], respectively. We magnified the oscillation amplitudes of the vortices centers by 2 times to a better demonstration.}
\label{FigureS5}
\end{figure}

\begin{figure}[ptbh]
\begin{centering}
\includegraphics[width=0.48\textwidth]{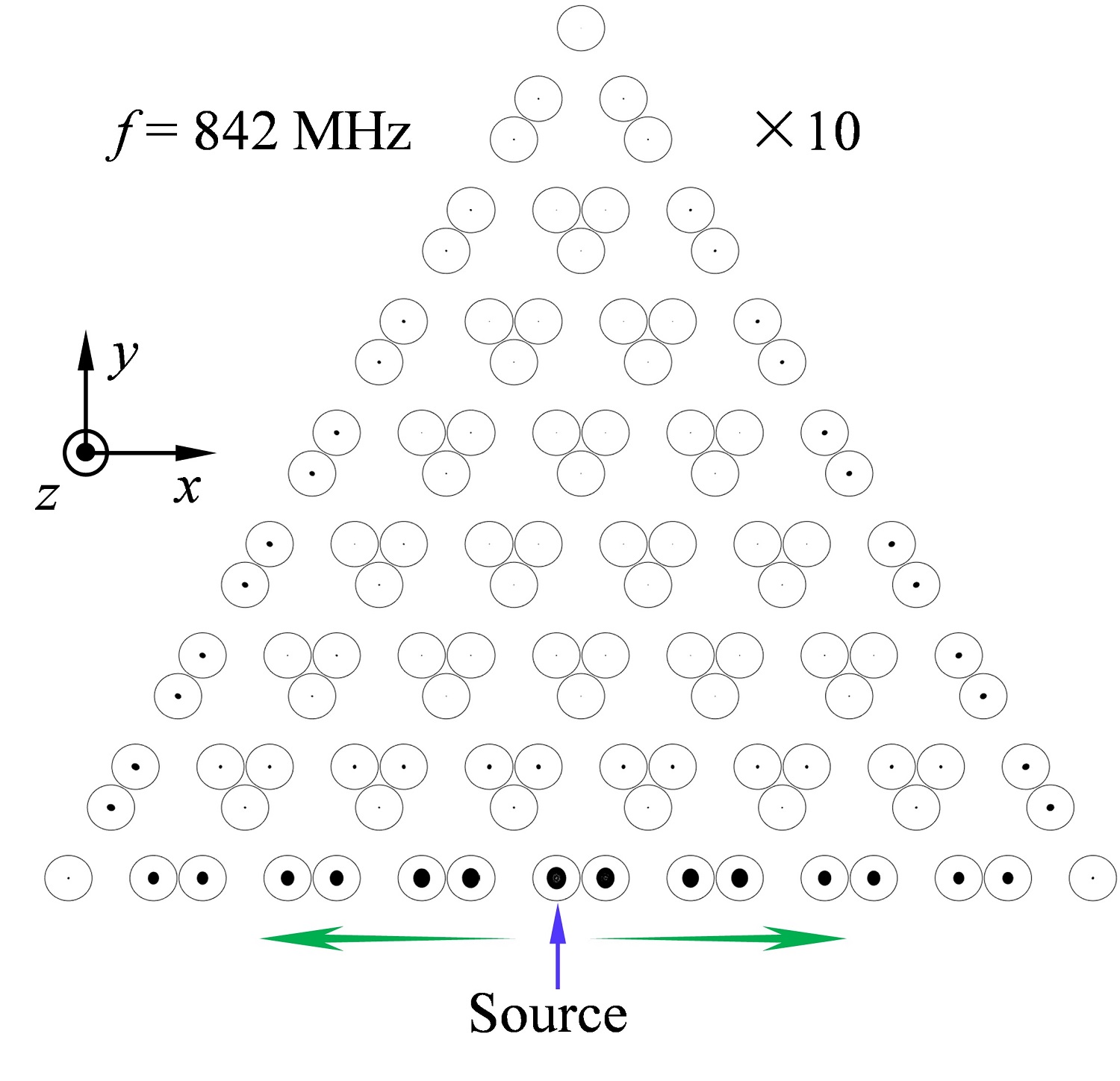}
\par\end{centering}
\caption{\textbf{Tamm-Shockley edge state:} Snapshot of the propagation of edge state with the frequency $f=842$ MHz at $t=100$ ns. Oscillation amplitudes of vortices centers have been magnified 10 times for better illustrations.}
\label{FigureS6}
\end{figure}

\begin{figure}[ptbh]
\begin{centering}
\includegraphics[width=0.48\textwidth]{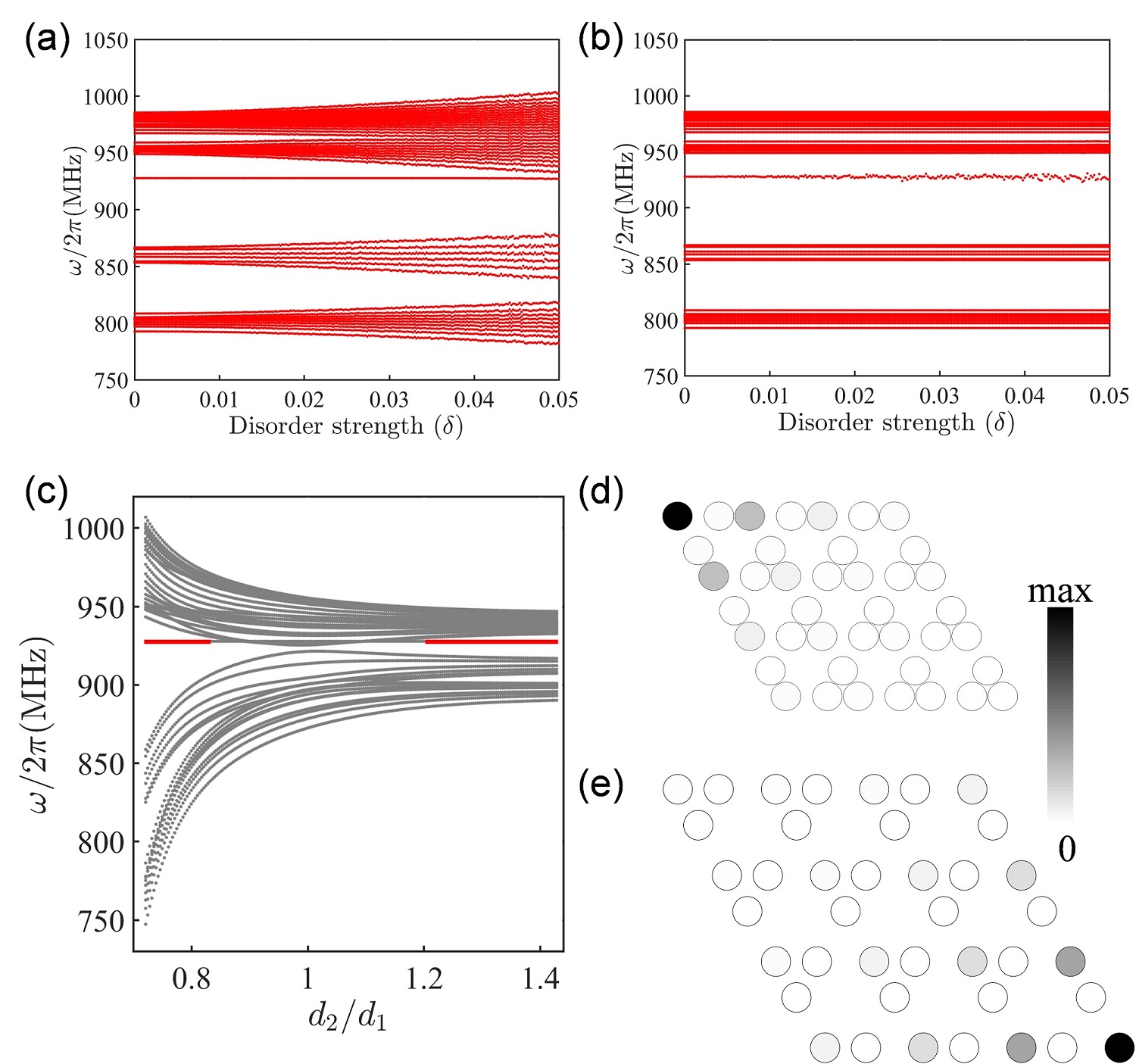}
\par\end{centering}
\caption{\textbf{The robustness of the corner states and the phase diagram for parallelogram-shape:} The eigenfrequencies of the parallelogram-shape breathing kagome lattice of vortices under the different disorder strength, (\textbf{a}) disorders in the bulk and (\textbf{b}) disorder at the position of one acute angle. (\textbf{c}) The eigenfrequencies of system for different values $d_{2}/d_{1}$ with $d_{1}=2.8r$. Two red line segments represent the region where the corner states exist. The spatial distribution of vortices oscillation for the corner states when (\textbf{d}) $d_{1}/d_{2}=1.35$ and (\textbf{e}) $d_{2}/d_{1}=1.4$. }
\label{FigureS7}
\end{figure}

\subsection*{\bf{Supplementary Note 4}}
\noindent The effect of disorders in the bulk on eigenfrequencies of triangle-shape system have been disscussed in the main text. For comparison, Supplementary Fig. \ref{FigureS4}(c) shows the results when disorders are located at three corners, we can seen that the frequencies of the corner states have great changes. Besides, the larger the strength of disorders, the bigger the frequency difference of the three corner states. To have a better understanding on the role of disorder on corner states, we plot the eigenfrequencies of system for disorders in the bulk [Supplementary Fig. \ref{FigureS4}(a)] and at corners [Supplementary Fig. \ref{FigureS4}(d)] with $\delta=0.01$. Supplementary Fig. \ref{FigureS4}(b) [Supplementary Figs. \ref{FigureS4}(e1)-\ref{FigureS4}(e3)] shows the spatial distribution of vortex gyrations for corner states as indicated in Supplementary Fig. \ref{FigureS4}(a) [Supplementary Fig. \ref{FigureS4}(d)]. The difference of frequencies for corner states in Supplementary Fig. \ref{FigureS4}(d) originates from the destruction of rotational symmetry by disorders.

In addition, we investigate the behavior of corner states under artificial lattice defects. Supplementary Figs. \ref{FigureS5}(a), \ref{FigureS5}(c), and \ref{FigureS5}(e) show the schematics for different type of defects. Then we apply a sinusoidal field $\textbf{h}(t)=h_0\sin(2\pi ft)\hat{x}$ with $h_0=0.1$ mT and $f=940$ MHz to the whole system. Supplementary Figs. \ref{FigureS5}(b), \ref{FigureS5}(d), and \ref{FigureS5}(f) show the spatial distribution of oscillation amplitude with different defect types. Micromagnetic simulations are run for 100 ns. One can clearly see that the corner states are robust for those defects. Interestingly, some defects may introduce new corners and corner states, see Supplementary Fig. \ref{FigureS5}(d). But corner states always emerge at acuted-angle corners.

\subsection*{\bf{Supplementary Note 5}}
\noindent In previous sections, we have shown that there exist edge states along three boundaries. Here, we confirm that the observed edge modes are Tamm-Shockley type instead of topological or chiral. We simulate the dynamics of vortex lattice by a sinusoidal field $\textbf{h}(t)=h_0\sin(2\pi ft)\hat{x}$ with $h_0=0.1$ mT, $f=842$ MHz applied on one disk at the edge, indicated by the blue arrow in Supplementary Fig. \ref{FigureS6}, with the same parameters as those in the main text. Then we draw the spatial distribution of oscillation amplitude, it can be seen clearly that the propagation of vortex oscillations is bidirectional. This nonchiral mode can be simply explained in terms of the Tamm-Shockley mechanism \cite{Tamm1932,Shockley1939,Li2018PRB}.

\subsection*{\bf{Supplementary Note 6}}
\noindent At last, we study the robustness of the corner states and the phase diagram for parallelogram-shape lattices. Supplementary Figs. \ref{FigureS7}(a) and \ref{FigureS7}(b) show the eigenfrequencies of system under different disorders localized in the bulk and the corner, respectively. Similar to the triangle-shape case, the corner state is robust against disorders in the bulk. When disorders are placed at the position of the acute angle, the corner state always exist and the oscillations of vortices are also localized at this acute angle, while the frequency of corner state is modified. The phase diagram of the clean system (with no disorders) is plotted in Supplementary Fig. \ref{FigureS7}(c). Unlike the triangular case, with the increasing of $d_{2}/d_{1}$, there exist two regions where the corner states emerge, as indicated by the red line segments. In the calculations, we have fixed $d_{1}=2.8r$. The spatial distribution of corner states oscillations is plotted in Supplementary Figs. \ref{FigureS7}(d) and \ref{FigureS7}(e). We can clearly see the oscillation of vortices localized at two different acute angles in two different regions, i.e., $d_{1}/d_{2}\geqslant1.2$ and $d_{2}/d_{1}\geqslant1.2$.


\begin{thebibliography}{99}

\bibitem {Hasan2010}Hasan, M. Z. \& Kane, C. L. Colloquium: Topological insulators. \textit{Rev. Mod. Phys.} \textbf{82}, 3045-3067 (2010).

\bibitem {Qi2011}Qi, X. L. \& Zhang, S. C. Topological insulators and superconductors. \textit{Rev. Mod. Phys.} \textbf{83}, 1057-1110 (2011).

\bibitem {Hsieh2009}Hsieh, D. \textit{et al.} A tunable topological insulator in the spin helical Dirac transport regime. \textit{Nature} \textbf{460}, 1101-1105 (2009).

\bibitem {Pribiag2015}Pribiag, V. S. \textit{et al.} Edge-mode superconductivity in a two-dimensional topological insulator. \textit{Nat. Nanotech.} \textbf{10}, 593-597 (2015).

\bibitem {Benalcazar2017}Benalcazar, W. A., Bernevig, B. A. \& Hughes, T. L. Quantized electric multipole insulators. \textit{Science} \textbf{357}, 61-66 (2017).

\bibitem {Bernevig2017}Benalcazar, W. A., Bernevig, B. A. \& Hughes, T. L. Electric multipole moments, topological multipole moment pumping, and chiral hinge states in crystalline insulators, \textit{Phys. Rev. B} \textbf{96}, 245115 (2017).

\bibitem {Song2017}Song, Z., Fang, Z. \& Fang, C. (\textit{d}-2)-Dimensional Edge States of Rotation Symmetry Protected Topological States. \textit{Phys. Rev. Lett.} \textbf{119}, 246402 (2017).

\bibitem {Langbehn2017}Langbehn, J., Peng, Y., Trifunovic, L., von Oppen, F. \& Brouwer, P. W. Reflection-Symmetric Second-Order Topological Insulators and Superconductors. \textit{Phys. Rev. Lett.} \textbf{119}, 246401 (2017).

\bibitem {Schindler2018}Schindler, F. \textit{et al.} Higher-order topological insulators. \textit{Sci. Adv.} \textbf{4}, eaat0346 (2018).

\bibitem {Ezawa2018}Ezawa, M. Higher-Order Topological Insulators and Semimetals on the Breathing Kagome and Pyrochlore Lattices. \textit{Phys. Rev. Lett.} \textbf{120}, 026801 (2018).

\bibitem {Ezawa2018_2}Ezawa, M. Higher-order topological electric circuits and topological corner resonance on the breathing kagome and pyrochlore lattices. \textit{Phys. Rev. B} \textbf{98}, 201402(R) (2018).

\bibitem {Xie2018}Xie, B. Y. \textit{et al.} Second-order photonic topological insulator with corner states. \textit{Phys. Rev. B} \textbf{98}, 205147 (2018).

\bibitem {Serra2018}Serra-Garcia, M. \textit{et al.} Observation of a phononic quadrupole topological insulator. \textit{Nature} \textbf{555}, 342-345 (2018).

\bibitem {Peterson2018}Peterson, C. W., Benalcazar, W. A., Hughes, T. L. \& Bahl, G. A quantized microwave quadrupole insulator with topologically protected corner states, \textit{Nature} \textbf{555}, 346-350 (2018).

\bibitem {Xue2019}Xue, H., Yang, Y., Gao, F., Chong, Y. \& Zhang, B. Acoustic higher-order topological insulator on a kagome lattice. \textit{Nat. Mater.} \textbf{18}, 108-112 (2019).

\bibitem {Ni2019}Ni, X., Weiner, M., Al, A. \& Khanikaev, A. B. Observation of higher-order topological acoustic states protected by generalized chiral symmetry. \textit{Nat. Mater.} \textbf{18}, 113-120 (2019).

\bibitem {Hassan2018}Hassan, A. E. \textit{et al.} Corner states of light in photonic waveguides. Preprint at https://arxiv.org/abs/1812.08185 (2018).

\bibitem {Mittal2018}Mittal, S. \textit{et al.} Photonic quadrupole topological phases. Preprint at https://arxiv.org/abs/1812.09304 (2018).

\bibitem {Yang2019}Xue, H. \textit{et al.} Realization of an acoustic third-order topological insulator. \textit{Phys. Rev. Lett.} \textbf{122}, 244301 (2019).

\bibitem {Imhof2018}Imhof, S. \textit{et al.} Topolectrical-circuit realization of topological corner modes. \textit{Nat. Phys.} \textbf{14}, 925-929 (2018).

\bibitem {Serra2019}Serra-Garcia, M., S\"{u}sstrunk, R. \& Huber, S. D. Observation of quadrupole transitions and edge mode topology in an LC circuit network. \textit{Phys. Rev. B} \textbf{99}, 020304(R) (2019).

\bibitem{LFZhang2013}Zhang, L., Ren, J., Wang, J. S. \& Li, B. Topological magnon insulator in insulating ferromagnet. \textit{Phys. Rev. B} \textbf{87}, 144101 (2013).

\bibitem{Shindou2013}Shindou, R. \textit{et al.} Chiral spin-wave edge modes in dipolar magnetic thin films. \textit{Phys. Rev. B} \textbf{87}, 174402 (2013).

\bibitem{Mook2014}Mook, A., Henk, J. \& Mertig, I. Edge states in topological magnon insulators. \textit{Phys. Rev. B} 90, 024412 (2014).

\bibitem{XSWang2017}Wang, X. S., Su, Y. \& Wang, X. R. Topologically protected unidirectional edge spin waves and beam splitter. \textit{Phys. Rev. B} \textbf{95}, 014435 (2017).

\bibitem{Chernyshev2016}Chernyshev, A. L. \& Maksimov, P. A. Damped Topological Magnons in the Kagome-Lattice Ferromagnets. \textit{Phys. Rev. Lett.} \textbf{117}, 187203 (2016).

\bibitem{Mook2016}Mook, A., Henk, J. \& Mertig, I. Tunable Magnon Weyl Points in Ferromagnetic Pyrochlores. \textit{Phys. Rev. Lett.} \textbf{117}, 157204 (2016).

\bibitem{Su2017}Su, Y., Wang, X. S. \& Wang, X. R. Magnonic Weyl semimetal and chiral anomaly in pyrochlore ferromagnets. \textit{Phys. Rev. B} \textbf{95}, 224403 (2017).

\bibitem{Su2017S}Su, Y. \& Wang, X. R. Chiral anomaly of Weyl magnons in stacked honeycomb ferromagnets. \textit{Phys. Rev. B} \textbf{96}, (2017).

\bibitem {Wachowiak2002}Wachowiak, A. \textit{et al.} Direct Observation of Internal Spin Structure of Magnetic Vortex Cores. \textit{Science} \textbf{298}, 577-580 (2002).

\bibitem {Yamada2007}Yamada, K. \textit{et al.} Electrical switching of the vortex core in a magnetic disk. \textit{Nat. Mater.} \textbf{6}, 270-273 (2007).

\bibitem {Bakaul2011}Bakaul, S. R., Lin, W. \& Wu, T. Evolution of magnetic bubble domains in manganite films. \textit{Appl. Phys. Lett.} \textbf{99}, 042503 (2011).

\bibitem {Petit2015}Petit, D., Seem, P. R., Tillette, M., Mansell, R. \& Cowburn, R. P. Two-dimensional control of field-driven magnetic bubble movement using Dzyaloshinskii-Moriya interactions. \textit{Appl. Phys. Lett.} \textbf{106}, 022402 (2015).

\bibitem {Muhlbauer2009}M\"{u}hlbauer, S. \textit{et al.} Skyrmion lattice in a chiral magnet. \textit{Science} \textbf{323}, 915-919 (2009).

\bibitem {Jiang2015}Jiang, W. \textit{et al.} Blowing magnetic skyrmion bubbles. \textit{Science} \textbf{349}, 283-286 (2015).

\bibitem{Yang2018PRB}Yang, H. \textit{et al.} Twisted skyrmions at domain boundaries and the method of image skyrmions, \textit{Phys. Rev. B} \textbf{98}, 014433 (2018).

\bibitem {Yang2018PRL}Yang, H., Wang, C., Yu, T., Cao, Y. \& Yan, P. Antiferromagnetism Emerging in a Ferromagnet with Gain. \textit{Phys. Rev. Lett.} \textbf{121}, 197201 (2018).

\bibitem {Pribiag2007}Pribiag, V. S. \textit{et al.} Magnetic vortex oscillator driven by d.c. spin-polarized current. \textit{Nat. Phys.} \textbf{3}, 498-503 (2007).

\bibitem {Finocchio2016}Finocchio, G., B\"{u}ttner, F., Tomasello, R., Carpentieri, M. \& Kl\"{a}ui, M. Magnetic skyrmions: from fundamental to applications. \textit{J. Phys. D: Appl. Phys.} \textbf{49}, 423001 (2016).

\bibitem {Han2013}Han, D. \textit{et al.} Wave modes of collective vortex gyration in dipolar-coupled-dot-array magnonic crystals. \textit{Sci. Rep.} \textbf{3}, 2262 (2013).

\bibitem {Yang2017}Kim, J. \textit{et al.} Coupled gyration modes in one-dimensional skyrmion arrays in thin-film nanostrips as new type of information carrier. \textit{Sci. Rep.} \textbf{7}, 45185 (2017).

\bibitem {Mruczkiewicz2016}Mruczkiewicz, M., Gruszecki, P., Zelent, M. \& Krawczyk, M. Collective dynamical skyrmion excitations in a magnonic crystal. \textit{Phys. Rev. B} \textbf{93}, 174429 (2016).

\bibitem {Behncke2015}Behncke, C., H\"{a}nze, M., Adolff, C. F., Weigand, M. \& Meier, G. Band structure engineering of two-dimensional magnonic vortex crystals. \textit{Phys. Rev. B} \textbf{91}, 224417 (2015).
%
\bibitem {Adolff2016}H\"{a}nze, M. \textit{et al.} Collective modes in three-dimensional magnonic vortex crystals. \textit{Sci. Rep.} \textbf{6}, 22402 (2016).

\bibitem {Kruglyak2010}Kruglyak, V. V. \textit{et al.} Imaging Collective Magnonic Modes in 2D Arrays of Magnetic Nanoelements. \textit{Phys. Rev. Lett.} \textbf{104}, 027201 (2010).

\bibitem {Tacchi2011}Tacchi, S. \textit{et al.} Band Diagram of Spin Waves in a Two-Dimensional Magnonic Crystal. \textit{Phys. Rev. Lett.} \textbf{107}, 127204 (2011).

\bibitem {Krawczyk2014}Krawczyk, M. \& Grundler, D. Review and prospects of magnonic crystals and devices with reprogrammable band structure. \textit{J. Phys.: Condens. Matter} \textbf{26}, 123202 (2014).

\bibitem {Li2018PRB}Li, Z.-X., Wang, C., Cao, Y. \& Yan, P. Edge states in a two-dimensional honeycomb lattice of massive magnetic skyrmions. \textit{Phys. Rev. B} \textbf{98}, 180407(R) (2018).

\bibitem {Kim2017}Kim, S. K. \& Tserkovnyak, Y. Chiral Edge Mode in the Coupled Dynamics of Magnetic Solitons in a Honeycomb Lattice. \textit{Phys. Rev. Lett.} \textbf{119}, 077204 (2017).

\bibitem {Makhfudz2012}Makhfudz, I., Kr\"{u}ger, B. \& Tchernyshyov, O. Inertia and Chiral Edge Modes of a Skyrmion Magnetic Bubble. \textit{Phys. Rev. Lett.} \textbf{109}, 217201 (2012).

\bibitem {Yang2018OE}Yang, W., Yang, H., Cao, Y. \& Yan, P. Photonic orbital angular momentum transfer and magnetic skyrmion rotation. \textit{Opt. Express} \textbf{26}, 8778 (2018).

\bibitem {Buttner2015}B\"{u}ttner, F. \textit{et al.} Dynamics and inertia of skyrmionic spin structures. \textit{Nat. Phys.} \textbf{11}, 225-228 (2015).

\bibitem {Mertens1997}Mertens, F. G., Schnitzer, H. J. \& Bishop, A. R. Hierarchy of equations of motion for nonlinear coherent excitations applied to magnetic vortices. \textit{Phys. Rev. B} \textbf{56}, 2510 (1997).

\bibitem {Ivanov2010}Ivanov, B. A. \textit{et al.} Non-Newtonian dynamics of the fast motion of a magnetic vortex. \textit{JETP Lett.} \textbf{91}, 178 (2010).

\bibitem {Cherepov2012}Cherepov, S. S. \textit{et al.} Core-Core Dynamics in Spin Vortex Pairs. \textit{Phys. Rev. Lett.} \textbf{109}, 097204 (2012).

\bibitem {Shibata2003}Shibata, J., Shigeto, K. \& Otani, Y. Dynamics of magnetostatically coupled vortices in magnetic nanodisks. \textit{Phys. Rev. B} \textbf{67}, 224404 (2003).

\bibitem {Shibata2004}Shibata, J. \& Otani, Y. Magnetic vortex dynamics in a two-dimensional square lattice of ferromagnetic nanodisks. \textit{Phys. Rev. B} \textbf{70}, 012404 (2004).

\bibitem {Lee2011}Lee, K. S., Jung, H., Han, D. S. \& Kim, S. K. Normal modes of coupled vortex gyration in two spatially separated magnetic nanodisks. \textit{J. Appl. Phys.} \textbf{110}, 113903 (2011).

\bibitem {Sukhostavets2013}Sukhostavets, O. V., Gonz\'{a}lez, J. \& Guslienko, K. Y. Multipole magnetostatic interactions and collective vortex excitations in dot pairs, chains, and two-dimensional arrays. \textit{Phys. Rev. B} \textbf{87}, 094402 (2013).

\bibitem {Sinnecker2014}Sinnecker, J. P., Vigo-Cotrina, H., Garcia, F., Novais, E. R. P. \& Guimar\~{a}es, A. P. Interaction between magnetic vortex cores in a pair of nonidentical nanodisks. \textit{J. Appl. Phys.} \textbf{115}, 203902 (2014).

\bibitem {Yoo2012}Yoo, M. W., Lee, J. \& Kim, S. K. Radial-spin-wave-mode-assisted vortex-core magnetization reversals. \textit{Appl. Phys. Lett.} \textbf{100}, 172413 (2012).

\bibitem {Velten2017}Velten, S. \textit{et al.} Vortex circulation patterns in planar microdisk arrays. \textit{Appl. Phys. Lett.} \textbf{110}, 262406 (2017).

\bibitem{Ivanov1998}Ivanov, B. A., Schnitzer, H. J., Mertens, F. G. \& Wysin, G. M. Magnon modes and magnon-vortex scattering in two-dimensional easy-plane ferromagnets. \textit{Phys. Rev. B} \textbf{58}, 8464 (1998).

\bibitem {Tamm1932}Tamm, I. \"{U}ber eine m\"{o}gliche Art der Elektronenbindung an Kristalloberfl\"{a}chen. \textit{Phys. Z. Sowjetunion.} \textbf{76}, 849 (1932).

\bibitem {Shockley1939}Shockley, W. On the surface states associated with a periodic potential. \textit{Phys. Rev.} \textbf{56}, 317 (1939).

\bibitem{Smith1993}King-Smith, R. D. \& Vanderbilt, D. Theory of polarization of crystalline solids. \textit{Phys. Rev. B} \textbf{47}, 1651(R) (1993).

\bibitem{Vanderbilt1993}Vanderbilt, D. \& King-Smith, R. D. Electric polarization as a bulk quantity and its relation to surface charge. \textit{Phys. Rev. B} \textbf{48}, 4442 (1993).

\bibitem {Vansteenkiste2014}Vansteenkiste, A. \textit{et al.} The design and verification of MuMax3. \textit{AIP Adv.} \textbf{4}, 107133 (2014).

\bibitem {Yoo2015}Yoo, M. W. \& Kim, S. K. Azimuthal-spin-wave-mode-driven vortex-core reversals. \textit{J. Appl. Phys.} \textbf{117}, 023904 (2015).

\end{thebibliography}
\end{document}